\documentclass[preprint,12pt]{myelsarticle}

\usepackage[normalem]{ulem}
\newcommand{\pp}[2]{\def\firstarg{#1}\def\secondarg{#2}\def\empty{}%
\sout{#1}%
\ifx\firstarg\empty\else\ifx\secondarg\empty\else{ }\fi\fi%
\uwave{#2}}

\usepackage[latin1]{inputenc}
\usepackage[T1]{fontenc}
\usepackage{ae}
\usepackage{latexsym,ifthen}
\usepackage{amsmath}
\usepackage{amssymb}
\usepackage[mathscr]{eucal}
\usepackage{graphicx}
\usepackage{enumerate,enumitem}
\usepackage{float}

\usepackage{tikz}
\usetikzlibrary{calc,positioning,arrows}

\newtheorem{theorem}{Theorem}
\newtheorem{lemma}{Lemma}
\newtheorem{proposition}{Proposition}
\newtheorem{corollary}{Corollary}
\newtheorem{observation}{Observation}

\newtheorem{definition}{Definition}
\newenvironment{proof}{\noindent {\bf Proof.\,}}{\qed}

\begin{document}

\begin{frontmatter}

\title{Characterization and recognition of some opposition and coalition graph classes
}

\author{Van Bang Le}
\ead{le@informatik.uni-rostock.de}
\address{Universit\"at Rostock, Institut f\"ur Informatik, Rostock, Germany}

\author{Thomas Podelleck}
\ead{thomas.podelleck@gmail.com}
\address{Universit\"at Rostock, Institut f\"ur Informatik, Rostock, Germany}

\begin{abstract}
A graph is an opposition graph, respectively, a coalition graph, if it admits an acyclic
orientation which puts the two end-edges of every chordless $4$-vertex path in opposition,
respectively, in the same direction. Opposition and coalition graphs have been introduced
and investigated in connection to perfect graphs.
Recognizing and characterizing opposition and coalition graphs are long-standing open problems.
This paper gives characterizations for opposition graphs and coalition graphs on
some restricted graph classes. Implicit in our arguments are polynomial time recognition
algorithms for these graphs. We also give a good characterization for the so-called
generalized opposition graphs.
\end{abstract}


\end{frontmatter}

\section{Introduction and preliminaries}
Chv\'atal \cite{Chvatal1984} proposed to call a linear order $<$ on the vertex set of an
undirected graph $G$ {\it perfect\/} if the greedy coloring algorithm applied to each induced
subgraph $H$ of $G$ gives an optimal coloring of $H$: Consider the vertices of $H$ sequentially
by following the order $<$ and assign to each vertex $v$ the smallest color not used on any
neighbor $u$ of $v$, $u<v$. A graph is {\it perfectly orderable\/} if it admits a perfect order.
Chv\'atal proved that $<$ is a perfect order if and only if there is no chordless path with
four vertices $a,b,c,d$ and three edges $ab, bc, cd$ (written $P_4\, abcd$) with $a<b$ and $d<c$.
He also proved that perfectly orderable graphs are perfect.\footnote{A graph is {\it perfect\/}
if the chromatic number and the clique number are equal in every induced subgraph.}
The class of perfectly orderable graphs properly contains many important, well-known classes of
perfect graphs such as chordal graphs and comparability graphs. Perfectly orderable graphs have
been extensively studied in the literature;
see Ho\`ang's comprehensive survey~\cite{Hoang2001} for more information.

\begin{figure}[H]
\begin{center}
\begin{tikzpicture}[scale=.6]
\tikzstyle{vertex}=[draw,circle,inner sep=1.5pt]
\node[vertex] (1) at (0,1)   {};
\node[vertex] (2) at (1.5,1) {};
\node[vertex] (3) at (3,1)   {};
\node[vertex] (4) at (4.5,1) {};

\draw[->,>=stealth'] (1) -- (2); 
\draw[->,>=stealth'] (2) -- (3);
\draw[->,>=stealth'] (4) -- (3);

\draw(2.25,0.2) node {type $0$};
\end{tikzpicture}
\end{center}

\begin{center}
\begin{tikzpicture}[scale=.6]
\tikzstyle{vertex}=[draw,circle,inner sep=1.5pt]
\node[vertex] (1) at (0,1)   {};
\node[vertex] (2) at (1.5,1) {};
\node[vertex] (3) at (3,1)   {};
\node[vertex] (4) at (4.5,1) {};

\draw[->,>=stealth'] (2) -- (1); 
\draw[->,>=stealth'] (2) -- (3); 
\draw[->,>=stealth'] (3) -- (4); 

\draw(2.25,0.2) node {type $1$};
\end{tikzpicture}
\qquad
\begin{tikzpicture}[scale=.6]
\tikzstyle{vertex}=[draw,circle,inner sep=1.5pt]
\node[vertex] (1) at (0,1)   {};
\node[vertex] (2) at (1.5,1) {};
\node[vertex] (3) at (3,1)   {};
\node[vertex] (4) at (4.5,1) {};

\draw[->,>=stealth'] (1) -- (2); 
\draw[->,>=stealth'] (2) -- (3); 
\draw[->,>=stealth'] (3) -- (4); 

\draw(2.25,0.2) node {type $2$};
\end{tikzpicture}
\qquad
\begin{tikzpicture}[scale=.6]
\tikzstyle{vertex}=[draw,circle,inner sep=1.5pt]
\node[vertex] (1) at (0,1)   {};
\node[vertex] (2) at (1.5,1) {};
\node[vertex] (3) at (3,1)   {};
\node[vertex] (4) at (4.5,1) {};

\draw[->,>=stealth'] (1) -- (2); 
\draw[->,>=stealth'] (3) -- (2); 
\draw[->,>=stealth'] (3) -- (4); 

\draw(2.25,0.2) node {type $3$};
\end{tikzpicture}
\end{center}
\caption{Four types of oriented $P_4$.}\label{fig:1}
\end{figure}
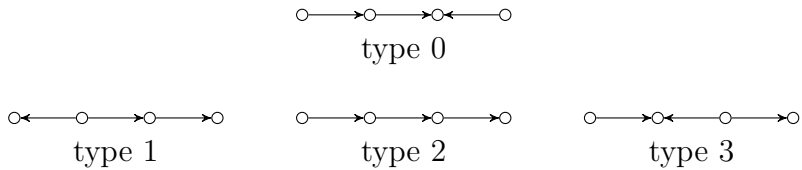

Recognizing perfectly orderable graphs is NP-complete \cite{MidPfe} (see also \cite{Hoang1996b}).
Also, no characterization of perfectly orderable graphs by forbidden induced subgraphs is known.
These facts have motivated researchers to study subclasses of perfectly orderable graphs;
see, e.g.,~\cite{EJSS,Hoang2001,HoangReed} and the literature given there.
Observe that a linear order $<$ corresponds to an acyclic orientation by directing the
edge $xy$ from $x$ to $y$ if and only if $x<y$.
Thus, a graph is perfectly orderable if and only if it admits an acyclic orientation such
that the orientation of no chordless path $P_4$ is of type $0$ (equivalently, the orientation of every $P_4$ is of type $1, 2$, or~$3$); see Figure~\ref{fig:1}.

One of the natural subclass of perfectly orderable graphs for which the recognition
complexity, as well as an induced subgraph characterization are still unknown is the following
(cf.~\cite{Hoang2001,HoangReed}).

\begin{definition}\label{def:coali}
A graph is a \emph{coalition graph} if it admits an acyclic orientation such that every
induced $P_4\, abcd$ has the end-edges $ab$ and $cd$ oriented in the \lq same way\rq, that is,
every oriented $P_4$ is of type $2$ or~$3$. Such an orientation is called a \emph{coalition orientation}.
\end{definition}

Equivalently, a graph is a coalition graph if it admits a linear order $<$ on its vertex set
such that every induced $P_4\, abcd$ has $a<b$ if and only if $c<d$.
In \cite{Hoang2001}, coalition graphs are called one-in-one-out graphs.
Examples of coalition graphs include comparability graphs, hence all bipartite graphs.

A related graph class has been introduced by Olariu in~\cite{Olariu}:

\begin{definition}\label{def:oppo}
A graph is an \emph{opposition graph} if it admits an acyclic orientation such that every
induced $P_4\, abcd$ has the end-edges $ab$ and $cd$
oriented \lq in opposition\rq, that is, every oriented $P_4$ is of type $0$ or~$1$.
Such an orientation is called an \emph{opposition orientation}.
\end{definition}

Equivalently, a graph is an opposition graph if it admits a linear order $<$ on its vertex set such that every $P_4\, abcd$ has $a<b$ if and only if $d<c$.
Olariu~\cite{Olariu} proved that opposition graphs are perfect.
He also conjectures (\cite{Olariu1997}) that not all opposition graphs are perfectly orderable. Examples of opposition graphs include all split graphs.
The recognition and characterization problems for opposition graphs are still open.

Coalition graphs and opposition graphs have been studied in the past from the combinatorial and algorithmic point of view. The characterization and recognition problems for these graphs have been solved for a few special graph classes so far.
A natural subclass of opposition graphs consists of those admitting an acyclic orientation in which every $P_4$ is oriented as type 1 (equivalently, every $P_4$ is oriented as type 0).
These are called \emph{bipolarizable graphs}, and have been characterized by (infinitely many) forbidden induced subgraphs in~\cite{Hertz,HoangReed}, and have been recognized using $O(n)$ adjacency matrix multiplications and thus in $O(n^{3.376})$ time in~\cite{EJSS}, and in $O(nm)$ time in~\cite{NP}; $n$ is the vertex number and $m$ is the edge number of the input graph. Another subclass of opposition graphs are the so-called Welsh-Powell opposition graphs; see~\cite{OlaRan,LinOla,NikPal} for more information.

In a recent paper~\cite{Le2013}, bipartite opposition graphs have been characterized by
(infinitely many) forbidden induced subgraphs, and have been recognized in linear time.
This paper gives also characterizations for complements of bipartite graphs that are coalition or opposition graphs.
It turns out that co-bipartite coalition graphs and co-bipartite opposition graphs coincide, and they are exactly the complements of bipartite permutation graphs, hence can be recognized in linear time. There is also a characterization of co-bipartite coalition/opposition graphs in terms of their bi-matrices. This characterization is similar to the one of co-bipartite perfectly orderable graphs given by Chv\'{a}tal in~\cite{Chvatal1993}, which has a close connection to a theorem in mathematical programming.

We first address in Section~\ref{sec:genoppo} the so-called generalized opposition graphs
introduced by Chv\'{a}tal; these graphs are obtained when the condition \lq acyclic\rq\ in Definition~\ref{def:oppo} is dropped. This concept turns out to be useful when considering opposition graphs in certain graph classes. We give a characterization for generalized opposition graphs in terms of an auxiliary graph, which leads to a polynomial time recognition algorithm.

In Section~\ref{sec:gemhouse} we extend the results in~\cite{Le2013} on bipartite opposition graphs to the larger class of $(\text{gem, house})$-free graphs. It turns out that inside this graph class, opposition graphs and generalized opposition graphs coincide, hence $(\text{gem, house})$-free opposition graphs can be recognized in polynomial time.

In Section~\ref{sec:dhoppo} we give a forbidden subgraph characterization for distance-hereditary opposition graphs, a subclass of $(\text{gem, house})$-free opposition gra\-p\-hs; this result includes the subgraph characterization for tree opposition graphs found in~\cite{Le2013}.

In Section~\ref{sec:dhcoali} we show that $(\text{gem, house, hole})$-free coalition graphs can be recognized in polynomial time by modifying the auxiliary graph for generalized opposition graphs. 
For the smaller class of distance-hereditary coalition graphs, we give a faster recognition algorithm by showing that they are indeed comparability graphs.
\paragraph{Definitions and notation}\, We consider only finite, simple, and undirected graphs.
For a graph $G$, the vertex set is denoted $V(G)$ and the edge set is denoted $E(G)$.
For a vertex $u$ of a graph $G$, the neighborhood of $u$ in $G$ is denoted $N_G(u)$ or simply $N(u)$ if the context is clear, and the degree of $u$ is $\deg(u)=|N(u)|$. Write $N[u]=N(u)\cup\{u\}$. 
For a set $U$ of vertices of a graph $G$, write $N(U)=\bigcup_{u\in U} N(u)\setminus U$ and $N[U]=N(U)\cup U$. The subgraph of $G$ induced by $U$ is denoted $G[U]$.
If $u$ is a vertex of a graph $G$, then $G-u$ is $G[V(G)\setminus \{ u\}]$.

For $\ell \ge 1$, let $P_\ell$ denote a chordless path with $\ell$ vertices and $\ell-1$
edges, and for $\ell \ge 3$, let $C_\ell$ denote a chordless cycle with $\ell$ vertices
and $\ell$ edges. 
We write $P_\ell\, u_1u_2\ldots u_\ell$ and $C_\ell\, u_1u_2\ldots u_\ell u_1$, meaning the chordless path with vertices $u_1, u_2,\ldots, u_\ell$ and edges $u_iu_{i+1}$, $1\le i<\ell$, respectively, the chordless cycle with vertices $u_1, u_2,\ldots, u_\ell$ and edges $u_iu_{i+1}$, $1\le i<\ell$, and $u_\ell u_1$. 
The edges $u_1u_2$ and $u_{\ell-1}u_\ell$ of the path $P_\ell$ ($\ell\ge 3$) are the {\it end-edges\/} and the other edges are the {\em mid-edges\/} of the path, while the vertices $u_1$ and $u_\ell$ are the {\em end-vertices\/} and the other vertices are the {\em mid-vertices\/} of the path. 
In this paper, all paths $P_\ell$ and all cycles $C_\ell$ will always be induced.

An {\it orientation\/} of an undirected graph $G$ is a directed graph $D(G)$ obtained from $G$ by replacing each edge $xy$ of $G$ by either $x \to y$ or $y\to x$ (but not both).
Recall that a graph is a comparability graph if it admits an acyclic orientation of its
edges such that whenever we have $a \to b$ and $b \to c$ we also have $a \to c$.
Equivalently, a graph is a comparability graph if it admits an acyclic orientation which
puts the two edges of every $P_3$ in opposition.
It follows that comparability graphs are coalition graphs.
Complements of bipartite/comparability graphs are {\it co-bipartite/co-comparability graphs\/}.

A cycle $C_k$ of length $k\ge 5$ is also called a {\it hole\/}, and the complement of the path $P_5$ is also called a {\it house\/}. A {\it gem\/} is obtained from the path $P_4$ by adding a vertex adjacent to all four vertices of the $P_4$. A {\it domino\/} is obtained from the cycle $C_6$ by adding a chord between two (distance $3$)-vertices.

A {\it chordal\/} graph has no induced $C_k$, $k\ge 4$. A {\it distance-hereditary\/} graph has no gem, no house, no domino, and no hole as induced subgraphs. 
Chordal distance-hereditary graphs are also called {\it ptolemaic\/} graphs. 
For more information on graph classes appearing in this paper, and for basic graph notation and definitions not given here, see, e.g.,~\cite{BLS,Spinrad}.

\section{Generalized opposition graphs}\label{sec:genoppo}

It was conjectured by Chv\'atal~\cite{Chvatal2000} and is implied by the strong perfect
graph theorem,\footnote{The strong perfect graph theorem (\cite{CRST}) states that graphs
without chordless cycles of odd length at least five and without complements of such a cycle are perfect.}
that if the term \lq acyclic\rq\ in Definition~\ref{def:oppo} of opposition graphs is dropped, the larger class is still a class of perfect graphs. Chv\'atal~\cite{Chvatal2000} proposed to call these graphs {\it generalized opposition graphs\/}.

\begin{definition}\label{def:genoppo}
A graph is a \emph{generalized opposition graph} if it admits an orientation such that every oriented $P_4$ is of type $0$ or $1$. Such an orientation is called a
\emph{generalized opposition orientation}.
\end{definition}

The co-bipartite graphs $\overline{C_{2k}}$, $k\ge 3$, are examples of generalized opposition graphs that are non-opposition graphs.

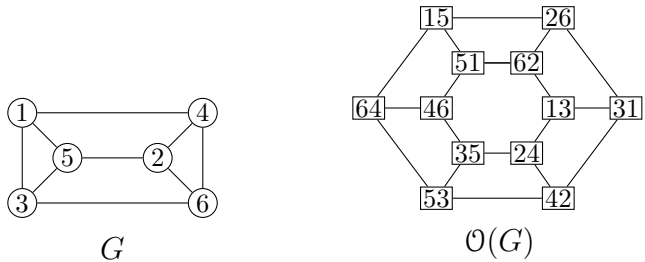
\begin{figure}[ht]
\begin{center}
\begin{tikzpicture}[scale=.6] 
\tikzstyle{vertex}=[draw,circle,inner sep=1pt]
\node[vertex] (1) at (0,3) {\footnotesize $1$};
\node[vertex] (2) at (3,2) {\footnotesize $2$};
\node[vertex] (3) at (0,1) {\footnotesize $3$};
\node[vertex] (4) at (4,3) {\footnotesize $4$};
\node[vertex] (5) at (1,2) {\footnotesize $5$};
\node[vertex] (6) at (4,1) {\footnotesize $6$};

\draw (1) -- (3) -- (5) -- (1);
\draw (2) -- (4) -- (6) -- (2);
\draw (1) -- (4); \draw (3) -- (6); \draw (5) -- (2);

\draw(2,0) node {$G$};
\end{tikzpicture}
\qquad\qquad
\begin{tikzpicture}[scale=.6] 
\tikzstyle{vertex}=[draw,rectangle,inner sep=1pt]

\node[vertex] (15) at (1.5,5)   {\footnotesize $15$};
\node[vertex] (26) at (4.2,5) {\footnotesize $26$};
\node[vertex] (31) at (5.7,3)   {\footnotesize $31$};
\node[vertex] (42) at (4.2,1) {\footnotesize $42$};
\node[vertex] (53) at (1.5,1)   {\footnotesize $53$};
\node[vertex] (64) at (0,3) {\footnotesize $64$};
\node[vertex] (51) at (2.2,4)   {\footnotesize $51$};
\node[vertex] (62) at (3.5,4) {\footnotesize $62$};
\node[vertex] (13) at (4.2,3)   {\footnotesize $13$};
\node[vertex] (24) at (3.5,2) {\footnotesize $24$};
\node[vertex] (35) at (2.2,2)   {\footnotesize $35$};
\node[vertex] (46) at (1.5,3) {\footnotesize $46$};

\draw (51) -- (15); \draw (62) -- (26);
\draw (15) -- (26); \draw (51) -- (62);
\draw (26) -- (31) -- (42) -- (53) -- (64) -- (15);
\draw (51) -- (62) -- (13) -- (24) -- (35) -- (46) -- (51);
\draw (13) -- (31);
\draw (24) -- (42); \draw (35) -- (53); \draw (46) -- (64);
\draw(2.9,0) node {$\mathscr{O}(G)$};
\end{tikzpicture}
\end{center}
\caption{Example of a graph $G$ and the auxiliary graph $\mathscr{O}(G)$; $xy$ stands for $(x,y)$.}\label{fig:genoppo}
\end{figure}

The main result of this section is a good characterization of generalized opposition
graphs in terms of an auxiliary graph defined below. Given a graph $G=(V,E)$, the
graph $\mathscr{O}(G)$ has
\begin{itemize}
 \item $\left\{(x,y)\mid \{x,y\} \text{ is an end-edge of an induced $P_4$ in $G$}\right\}$ as vertex set,
 \item two vertices $(x,y)$ and $(u,v)$ are adjacent in $\mathscr{O}(G)$ if $(x,y)=(v,u)$, or else
    $xyuv$\, or\, $uvxy$\, is an induced $P_4$ in $G$.
\end{itemize}
Notice that for each edge $\{x,y\}$ of $G$ that is an end-edge of an induced $P_4$ in $G$
there are two corresponding vertices in $\mathscr{O}(G)$, namely $(x,y)$ and $(y,x)$.
These two vertices are adjacent in $\mathscr{O}(G)$ and indicate the two possible orientations for the edge $\{x,y\}$. See Figure~\ref{fig:genoppo} for an example.
\begin{theorem}\label{thm:genoppo}
A graph $G$ is a generalized opposition graph if and only if $\mathscr{O}(G)$ is bipartite.
\end{theorem}
\proof Write $\mathscr{V}$ for the vertex set of $\mathscr{O}(G)$.

Suppose $G$ is an opposition graph, and consider an opposition orientation $D=D(G)$ of $G$.
Partition $\mathscr{V}$ into
$$
 A=\{(x,y)\in \mathscr{V}\mid x\to y\in D\}\,\, \text{and}\,\,  B=\{(x,y)\in \mathscr{V}\mid y\to x\in D\}.
$$
Then, by definition of adjacencies in $\mathscr{O}(G)$, $A$ and $B$ are independent sets in $\mathscr{O}(G)$, i.e., $\mathscr{O}(G)$ is bipartite.

Suppose $\mathscr{O}(G)$ is bipartite, and consider a bipartition $\mathscr{V}=A\cup B$ into
independent sets $A$ and $B$. We first orient the end-edges $\{x,y\}$ of $P_4$'s in $G$ according
to the bipartition: The edge $\{x,y\}$ gets the direction $x\to y$ if $(x,y)\in A$, otherwise $y\to x$.
Since $(x,y)$ and $(y,x)$ are adjacent in $\mathscr{O}(G)$, orienting the edge $\{x,y\}$ of $G$ in this way
is well-defined.
Now, extending this partial orientation $D(A)$ to the remaining edges of $G$ in an arbitrary way,
we obtain an orientation $D(G)$ of $G$, which is indeed a generalized opposition orientation of $G$.
\qed

\medskip
Since, given a graph $G$ with $m$ edges, $\mathscr{O}(G)$ can be constructed in $O(m^2)$ steps, we have:
\begin{corollary}\label{cor1}
Generalized opposition graphs with $m$ edges can be recognized in $O(m^2)$ time.
Moreover, a generalized opposition orientation of a generalized opposition graph can
be obtained in the same time.
\end{corollary}

Assume $\mathscr{O}(G)$ is bipartite and $\mathscr{V}=A\cup B$ is a bipartition of
$\mathscr{O}(G)$, and let $D(A)$ be the partial orientation on the end-edges of $P_4$'s in $G$
in the second part of the proof of Theorem~\ref{thm:genoppo} above. Then, if $D(A)$ is
{\it acyclic\/}, we can extend it to an {\it acyclic\/} orientation $D(G)$ of $G$. Hence we have:

\begin{proposition}\label{pro:oppo}
A graph $G$ is an opposition graph if and only if
\begin{itemize}
 \item[\em (i)] $\mathscr{O}(G)$ is bipartite, and
 \item[\em (ii)] there exists a bipartition $A$, $B$ of $\mathscr{O}(G)$ such that $D(A)$
   is acyclic.
\end{itemize}
\end{proposition}

As connected bipartite graphs have only one bipartition, and checking if a directed graph
is acyclic is doable in linear time, we obtain:
\begin{corollary}
Given $G$ such that $\mathscr{O}(G)$ is connected, it can be decided in linear time if $G$
is an opposition graph.
\end{corollary}
It would be interesting to discuss opposition graphs $G$ in the case of {\em disconnected\/} $\mathscr{O}(G)$.

\section{$(\text{gem, house})$-free opposition graphs}\label{sec:gemhouse}

Notice that opposition graphs are special cases of generalized opposition gra\-p\-hs. So, if $G$ is an opposition graph, then, by Theorem~\ref{thm:genoppo}, $\mathscr{O}(G)$ is bipartite.
The converse is, in general, not true; recall the example depicted in Figure~\ref{fig:genoppo}.
However, the converse does hold true for certain special graph classes, e.g., bipartite graphs as proved in~\cite{Le2013} recently.
In this section, we extend this result for $(\text{gem, house})$-free graphs.

\begin{theorem}\label{thm:gemhouse}
The following statements are equivalent for a $(\text{gem, house})$-free graph $G$:
\begin{itemize}
 \item[\em (i)] $G$ is an opposition graph;
 \item[\em (ii)] $G$ is a generalized opposition graph;
 \item[\em (iii)] $\mathscr{O}(G)$ is bipartite.
\end{itemize}
\end{theorem}
\proof
The implication (i) $\Rightarrow$ (ii) is obvious, (ii) $\Rightarrow$ (iii) is implied by Theorem~\ref{thm:genoppo}.
It remains to prove the implication (iii) $\Rightarrow$ (i). Suppose that $\mathscr{O}(G)$ is bipartite,
and $A$, $B$ is an arbitrary bipartition of $\mathscr{O}(G)$. In view of Proposition~\ref{pro:oppo},
we will show that $D(A)$ is acyclic. Recall that an edge $\{x,y\}$ of $G$ that
is an end-edge of an induced $P_4$ is oriented $x\to y$ in $D(A)$ if $(x,y)\in A$, otherwise
$y\to x\in D(A)$. Also recall that any induced $P_4\, xyuv$ in $G$ is \lq good\rq: it has $x\to y$ and
$v\to u$ in $D(A)$ or else $y\to x$ and $u\to v$ in $D(A)$.

Assume, by contradiction, that $C\, v_1v_2 \ldots v_qv_1$ is a directed cycle in $D(A)$.
Note that, since $v_i\to v_{i+1}\in D(A)$, by definition of $D(A)$, $(v_i, v_{i+1})\in A$ for
all $i$. Choose such a cycle $C$ with minimum length $q\ge 3$. Then,
\begin{equation}\label{eq:1}
\text{no chord of $C$ is an end-edge of an induced $P_4$,}
\end{equation}
otherwise there would be a shorter directed cycle in $D(A)$.
Let $P$ be an induced $P_4$ containing $\{v_1,v_2\}$ as an end-edge. Let, 
without loss of generality, $P$ be $v_1v_2ab$ for some vertices $a$ and $b$. Since $(v_1,v_2)\in A$, $(a,b)\in B$, and hence $b\to a\in D(A)$. 

We distinguish three cases; in each case we will arrive at a contradiction.

\smallskip
\noindent \textsc{Case 1:} $a$ and $b$ are outside $C$. 
In this case $\{v_q, b\}\not\in E(G)$, otherwise $v_q, v_1, v_2, a, b$
would induce a $C_5$ (note that $\mathscr{O}(C_5)$ contains a $C_5$), a gem or a house in $G$.
Hence $\{v_q,a\}\not\in E(G)$, otherwise $v_1v_qab$ would be a \lq bad\rq\ induced
$P_4$ with $v_q\to v_1\in D(A)$ but $b\to a\in D(A)$. 
It follows that $q > 3$ and $\{v_q,v_2\}\not\in E(G)$, otherwise $v_3v_2ab$ would be a bad $P_4$ (if $q=3$) or the chord $v_qv_2$ of $C$ (if $q>3$) would be an end-edge of the $P_4\, v_qv_2ab$, contradicting (\ref{eq:1}). Thus, $v_qv_1v_2a$ is an induced $P_4$, hence
$a\to v_2\in D(A)$.

Next, we show that $\{v_q,v_3\}\not\in E(G)$. Assume not. Then $\{a,v_3\}\in E(G)$, otherwise $av_2v_3v_q$ would be a bad $P_4$ (if $q=4$) or $\{v_3,v_q\}$ would be a chord of $C$ that is an end-edge of the induced $P_4\, av_2v_3v_q$. But then $v_1, v_2, v_3, v_q$ and $a$ induce a house (if $\{v_1,v_3\}\not\in E(G))$ or a gem (otherwise). Thus, $\{v_q,v_3\}\not\in E(G)$. Hence $\{v_1,v_3\}\in E(G)$ (otherwise $v_qv_1v_2v_3$ would be a bad $P_4$) and $\{b,v_3\}\not\in E(G)$ (otherwise $v_1,v_2,v_3,a,b$ would induce a house or a gem).

Now, if $\{a,v_3\}\not\in E(G)$, then $bav_2v_3$ is a bad $P_4$, a contradiction. If $\{a,v_3\}\in E(G)$, then $\{v_1,v_3\}$ is a chord of $C$ that is an end-edge of the induced $P_4\, v_1v_3ab$, a contradiction to (\ref{eq:1}). Case~1 is settled.

\smallskip
\noindent \textsc{Case 2:} $a=v_i$ for some $q> i\ge 3$.
Then, by (\ref{eq:1}), $b$ is outside $C$ or else $b=v_{i-1}$. (Note that $b\to v_i\in D(A)$, hence $b\not=v_{i+1}$.) 
First, $\{v_q, b\}\not\in E(G)$, otherwise $v_1, v_2, a, b, v_q$ would induce a gem, house or a $C_5$.
Hence $\{v_q,v_i\}\not\in E(G)$, otherwise $v_1v_qv_ib$ would be a bad $P_4$, and $\{v_q,v_2\}\in E(G)$, otherwise $v_qv_1v_2v_3$ would be a bad $P_4$ (if $i=3$) or the chord $\{v_2,v_i\}$ of $C$ would be an end-edge of the induced $P_4\, v_qv_1v_2v_i$.
But now the chord $\{v_q,v_2\}$ of $C$ is an end-edge of the $P_4\, v_qv_2v_ib$, contradicting (\ref{eq:1}).
Case~2 is settled.

\smallskip
\noindent \textsc{Case 3:} $a$ is outside $C$ and $b=v_i$ for some $q > i \ge 4$.
Note that $v_i\to a\in D(A)$.
First, $\{v_q,v_i\}\not\in E(G)$ (otherwise $v_q, v_1, v_2, a, v_i$ would induce a $C_5$, a house or a gem),
$\{v_q,a\}\not\in E(G)$ (otherwise $v_1v_qav_i$ would be a bad $P_4$), and $\{v_q,v_2\}\not\in E(G)$ (otherwise
the chord $\{v_q,v_2\}$ of $C$ would be an end-edge of the induced $P_4\, v_qv_2av_i$). Thus, $v_qv_1v_2a$ is
an induced $P_4$, hence $a\to v_2\in D(A)$.
But now $C'\, av_2v_3\ldots v_ia$ is a directed cycle in $D(A)$ of length $i<q$, contradicting the choice of $C$.

\smallskip
Case~3 is settled, and the proof of Theorem~\ref{thm:gemhouse} is complete.
\qed

\smallskip
Since bipartite graphs are $(\text{gem, house})$-free, Theorem~\ref{thm:gemhouse} implies:
\begin{corollary}[\cite{Le2013}]
A bipartite graph is an opposition graph if and only if it is a generalized opposition graph.
\end{corollary}

Moreover, Theorem~\ref{thm:gemhouse} and Corollary~\ref{cor1} imply the following result:
\begin{corollary}
Given a $(\text{gem, house})$-free graph $G$ with $m$ edges, it can be decided  in $O(m^2)$ time if $G$ is an opposition graph, and if so, an opposition orientation of $G$ can be obtained in the same time.
\end{corollary}

\section{Distance-hereditary opposition graphs}\label{sec:dhoppo}
We were not able to find a forbidden subgraph characterization for $(\text{gem}$, $\text{house})$-free opposition graphs.
This section gives such a characterization for distance-hereditary opposition graphs which form an important subclass of $(\text{gem}$, $\text{house})$-free opposition graphs.

\begin{figure}[H]
\begin{center}
\begin{tikzpicture}[scale=.6] 
\tikzstyle{vertex}=[draw,circle,inner sep=1.5pt]
\node[vertex] (1) at (0,1)  {};
\node[vertex] (2) at (1,1)  {};
\node[vertex] (3) at (2,1) [label=below:{\small $1$}] {};
\node[vertex] (4) at (3,1) [label=below:{\small $2$}] {};
\node[vertex] (7) at (6,1) {}; 
\node[vertex] (8) at (7,1) [label=below:{\small $2k$}] {};
\node[vertex] (k1) at (8,1) {};
\node[vertex] (k2) at (9,1) {};
\node[vertex] (a) at (2,2)  {};
\node[vertex] (b) at (7,2)  {};

\node (x) at (4,1) {};
\node (y) at (5.5,1) {};

\draw (1) -- (2) -- (3) -- (4) -- (x);
\draw[dashed, step=.1mm] (x) -- (y);
\draw (y) -- (7) -- (8) -- (k1) -- (k2);
\draw (a) -- (3);
\draw (b) -- (8);

\node[] (name) at (8,0.3) [label=right:{$\mathsf{T}_k$}] {};
\end{tikzpicture}
\end{center}
\begin{center}
\begin{tikzpicture}[scale=.6] 
\tikzstyle{vertex}=[draw,circle,inner sep=1.5pt]
\node[vertex] (1) at (0,1)  {};
\node[vertex] (2) at (1,1)  {};
\node[vertex] (3) at (2,1)  {};
\node[vertex] (4) at (3,1)  {};
\node[vertex] (5) at (1,2)  {};
\node[vertex] (6) at (2,2)  {};

\draw (1) -- (2) -- (3) -- (4);
\draw (2) -- (5) -- (6) -- (3);

\draw(1.5,0.3) node {$\mathsf{A}$};
\end{tikzpicture}
\qquad
\begin{tikzpicture}[scale=.6] 
\tikzstyle{vertex}=[draw,circle,inner sep=1.5pt]
\node[vertex] (1) at (0,1)  {};
\node[vertex] (2) at (1,1)  {};
\node[vertex] (3) at (2,1)  {};
\node[vertex] (4) at (3,1)  {};
\node[vertex] (5) at (4,1)  {};
\node[vertex] (6) at (5,1)  {};
\node[vertex] (7) at (2.5,2)  {};
\node[vertex] (8) at (2.5,3)  {};
\node[vertex] (9) at (2.5,4)  {};

\draw (1) -- (2) -- (3) -- (4) -- (5) -- (6);
\draw (3) -- (7) -- (8) -- (9); \draw (4) -- (7);

\draw(2.5,0.3) node {$\mathsf{G_1}$};
\end{tikzpicture}
\qquad
\begin{tikzpicture}[scale=.6] 
\tikzstyle{vertex}=[draw,circle,inner sep=1.5pt]
\node[vertex] (1) at (0,2)  {};
\node[vertex] (2) at (1,2)  {};
\node[vertex] (3) at (2,2)  {};
\node[vertex] (4) at (3,2)  {};
\node[vertex] (5) at (4,2)  {};
\node[vertex] (6) at (5,2)  {};
\node[vertex] (7) at (2.5,1)  {};
\node[vertex] (8) at (2.5,3)  {};
\node[vertex] (9) at (2.5,4)  {};

\draw (1) -- (2) -- (3) -- (4) -- (5) -- (6);
\draw (7) -- (3) -- (8) -- (9); \draw (7) -- (4) -- (8);

\draw(2.5,0.3) node {$\mathsf{G_2}$};
\end{tikzpicture}
\end{center}
\caption{Forbidden graphs for distance-hereditary (generalized) opposition graphs.}\label{fig:dhoppo}
\end{figure}
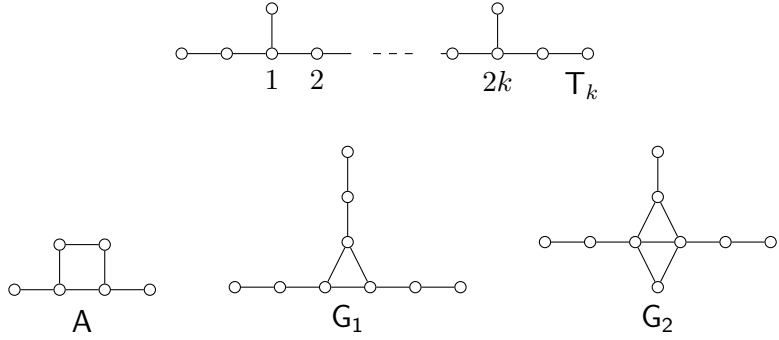

\begin{proposition}\label{pro:TkAG1G2}
The graphs $\mathsf{T}_k$ $(k\ge 1), \mathsf{A, G_1, G_2}$ depicted in Figure~$\ref{fig:dhoppo}$ are minimal non-opposition graphs.
\end{proposition}

\begin{theorem}\label{thm:dhoppo}
The following statements are equivalent for a distance-hereditary graph $G$:
\begin{itemize}
 \item[\em (i)] $G$ is an opposition graph;
 \item[\em (ii)] $G$ is a generalized opposition graph;
 \item[\em (iii)]  $\mathscr{O}(G)$ is bipartite;
 \item[\em (iv)] $G$ is $\mathsf{T}_k$-free $(k\ge 1)$ and $(\mathsf{A, G_1, G_2})$-free $(\text{see Figure}~\ref{fig:dhoppo})$.
\end{itemize}
\end{theorem}

The main and technical part in the proof of Theorem~\ref{thm:dhoppo} is the corresponding result for ptolemaic graphs (i.e., $\text{gem}$-free chordal graphs).

\begin{theorem}\label{thm:ptooppo}
The following statements are equivalent for a ptolemaic graph $G$:
\begin{itemize}
 \item[\em (i)] $G$ is an opposition graph;
 \item[\em (ii)] $G$ is a generalized opposition graph;
 \item[\em (iii)] $\mathscr{O}(G)$ is bipartite;
 \item[\em (iv)] $G$ is $\mathsf{T}_k$-free $(k\ge 1)$ and $(\mathsf{G_1, G_2})$-free $(\text{see Figure}~\ref{fig:dhoppo})$.
\end{itemize}
\end{theorem}

Theorem~\ref{thm:dhoppo} can be derived from Theorem~\ref{thm:ptooppo} as follows.
Two distinct vertices $u$ and $v$ of a graph $G$ are {\it twins\/} if $N_G(u)\setminus\{v\}=N_G(v)\setminus\{u\}$.
If $u, v$ are twins in $G$, then, clearly, $G$ is an opposition graph if and only if $G-u$ is an opposition graph. So, by induction, we may assume that the distance-hereditary graph $G$ in Theorem~\ref{thm:dhoppo} has no twins. Now, if $G$ has no $C_4$, then $G$ is chordal, and Theorem~\ref{thm:dhoppo} follows from Theorem~\ref{thm:ptooppo}.
If $G$ contains an induced $C_4$, then, as $G$ has no twins, it can be shown
(see also~\cite[Claim 3.5]{HoangReed}) that $G$ must contain an induced $\mathsf{A}$ or an induced house, or an induced domino. As $G$ is a distance-hereditary graph, none of these cases is possible, hence Theorem~\ref{thm:dhoppo}.

\medskip
\noindent
\textbf{Proof of Theorem~\ref{thm:ptooppo}.}\, 
(i) $\Rightarrow$ (ii) is obvious and (ii) $\Rightarrow$ (iii) follows from Theorem~\ref{thm:genoppo}. The implication (iii) $\Rightarrow$ (iv) follows by noting that the graphs $G$ depicted in Fig.~\ref{fig:dhoppo} have non-bipartite $\mathscr{O}(G)$. 

It remains to prove (iv) $\Rightarrow$ (i).
Let $G$ be a $\mathsf{T}_k$-free $(k\ge 1)$ and $(\mathsf{G_1, G_2})$-free ptolemaic graph. We have to show that $G$ is an opposition graph.

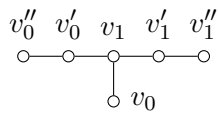
\begin{figure}[H]
\begin{center}
\begin{tikzpicture}[scale=.6] 
\tikzstyle{vertex}=[draw,circle,inner sep=1.5pt]
\node[vertex] (a) at (0,1)  [label=above:{\small $v_0''$}] {};
\node[vertex] (b) at (1,1)  [label=above:{\small $v_0'$}] {};
\node[vertex] (v1) at (2,1) [label=above:{\small $v_1$}] {};
\node[vertex] (v1a) at (3,1) [label=above:{\small $v_1'$}] {};
\node[vertex] (v1b) at (4,1) [label=above:{\small $v_1''$}] {};
\node[vertex] (c) at (2,0)  [label=right:{\small $v_0$}] {};

\draw (a) -- (b) -- (v1) -- (v1a) -- (v1b);
\draw (v1) -- (c);
\end{tikzpicture}
\end{center}
\caption{The graph $H_1$; vertex labeling only for the proof of Lemma~\ref{lem:final}.}\label{fig:H1}
\end{figure}

The proof is divided into three parts: $G$ is $P_5$-free (Lemma~\ref{lem:p5oppo}),
$G$ is $H_1$-free but contains an induced $P_5$ (Lemma~\ref{lem:h1oppo}), and
$G$ contains an induced $H_1$ (Lemma~\ref{lem:final}). See Figure~\ref{fig:H1} for the graph $H_1$.

\begin{lemma}\label{lem:p5oppo}
Every $P_5$-free ptolemaic graph is an opposition graph.
\end{lemma}
\proof A short proof uses the stronger result of \cite{Hertz}, which implies that every $P_5$-free ptolemaic graph is bipolarizable.\qed

\smallskip
Now, let $G$ be a ptolemaic graph, and let $w$ be a vertex of $G$.
For each integer $i\ge 0$, let $N_i(w)$ be the set of all vertices at distance exactly $i$ from $w$. Notice that $N_0(w)=\{w\}$. We will make use of the following facts.

\begin{observation}\label{obs}
Let $G$ be a ptolemaic graph. Let $i\ge 1$, and let $H$ be a connected component of $G[N_i(w)]$. Then $N(H)\cap N_{i-1}(w)$ is a clique and every vertex in $N(H)\cap N_{i-1}(w)$ is adjacent to all vertices in $H$.
\end{observation}
\proof We first show that 
\begin{equation}\label{eq:2}
\text{$N(H)\cap N_{i-1}(w)$ is a clique.}
\end{equation}
To see (\ref{eq:2}), suppose, by contradiction, $x\not=y$ are two non-adjacent vertices in $N(H)\cap N_{i-1}(w)$, and let $P$ and $Q$ be a shortest path in $G$ connecting $w$ and $x$, $w$ and $y$, respectively. Consider a neighbor $x'\in H$ of $x$ and a neighbor $y'\in H$ of $y$ with smallest distance in $H$, and let $R$ be a shortest path in $H$ connecting $x'$ and $y'$. Then $G[P\cup Q\cup R]$ contains an induced cycle (containing $R, x$ and $y$) of length at least four, contradicting the chordality of $G$, hence~(\ref{eq:2}).

Next we show, by induction on $i$, that
\begin{equation}\label{eq:3}
\text{every vertex in $N(H)\cap N_{i-1}(w)$ is adjacent to all vertices in $H$.}
\end{equation}
The case $i=1$ is obvious. Let $i\ge 2$, and suppose that some $x\in N(H)\cap N_{i-1}(w)$ is non-adjacent to some vertex in $H$. Then, since $H$ is connected, $x$ is adjacent to a vertex $u$ and non-adjacent to a vertex $v$ of $H$ such that $uv$ is an edge. Let $y\in N_{i-1}(w)$ be a neighbor of $v$. By~(\ref{eq:2}), $xy$ is an edge. Hence, by induction, there is some vertex $z\in N_{i-2}(w)$ adjacent to all vertices of the connected component in $G[N_{i-1}(w)]$ containing $x$ and $y$. But then $G[u,v,x,y,z]$ is a house or gem, a contradiction, hence~(\ref{eq:3}).\qed

\medskip
It follows from Observation~\ref{obs} that, for any $i\ge 0$, $G[N_i(w)]$ is $P_4$-free, otherwise $G$ would contain an induced gem. 
Let $P$ be an induced $P_4$ in $G$. If $P$ has three vertices in $N_i(w)$, then, by applying Observation~\ref{obs} to the connected components of $G[N_i(w)]$ containing vertices of $P$, we conclude that one end-vertex of $P$  must belong to $N_{i+1}(w)$. Similarly, if $P$ has exactly two vertices in $N_i(w)$, then (i) the mid-edge of $P$ is in $ N_i(w)$ and the end-vertices of $P$ are in $N_{i+1}(w)$, or (ii) an end-edge of $P$ is in $N_i(w)$, one mid-vertex of $P$ is in $N_{i+1}(w)$, and one end-vertex of $P$ is in $N_{i+2}(w)$, or (iii) one mid-vertex of $P$ is in $N_{i-1}(w)$ and one end-vertex of $P$ is in $N_{i+1}(w)$. 

\begin{figure}[ht]
\begin{center}
\begin{tikzpicture}[scale=.47]
\tikzstyle{vertex}=[draw,circle,inner sep=1.55pt] 

 \draw[step=1cm,gray!16,very thin] (-2.0cm,-1.0cm) grid (27.cm,9.0cm);

\draw (0,8) node {$N_i(w)$};
\draw (0,6) node {$N_{i+1}(w)$};
\draw (0,4) node {$N_{i+2}(w)$};
\draw (0,2) node {$N_{i+3}(w)$};
 
\node[vertex] (Aa) at (4,8) [label=right:$a$] {};
\node[vertex] (Ab) at (4,6) [label=right:$b$] {};
\node[vertex] (Ac) at (4,4) [label=right:$c$] {};
\node[vertex] (Ad) at (4,2) [label=right:$d$] {};
\draw[thick] (Aa) -- (Ab) -- (Ac) -- (Ad);
\draw[thick] (4,0) node {(A)};

\node[vertex] (Ba) at (7,6) [label=left:$a$] {};
\node[vertex] (Bb) at (7,8) [label=left:$b$] {};
\node[vertex] (Bc) at (9,8) [label=right:$c$] {};
\node[vertex] (Bd) at (9,6) [label=right:$d$] {};
\draw[thick] (Ba) -- (Bb) -- (Bc) -- (Bd);
\draw[thick] (8,0) node {(B)};

\node[vertex] (Ca) at (12,6) [label=left:$a$] {};
\node[vertex] (Cb) at (13,8) [label=right:$b$] {};
\node[vertex] (Cc) at (14,6) [label=right:$c$] {};
\node[vertex] (Cd) at (14,4) [label=right:$d$] {};
\draw[thick] (Ca) -- (Cb) -- (Cc) -- (Cd);
\draw[thick] (13,0) node {(C)};

\node[vertex] (Da) at (17,8) [label=left:$a$] {};
\node[vertex] (Db) at (19,8) [label=right:$b$] {};
\node[vertex] (Dc) at (19,6) [label=right:$c$] {};
\node[vertex] (Dd) at (19,4) [label=right:$d$] {};
\draw[thick] (Da) -- (Db) -- (Dc) -- (Dd);
\draw[thick] (18,0) node {(D)};

\node[vertex] (Ea) at (22,8) [label=above:$a$] {};
\node[vertex] (Eb) at (24,8) [label=above:$b$] {};
\node[vertex] (Ec) at (26,8) [label=above:$c$] {};
\node[vertex] (Ed) at (26,6) [label=right:$d$] {};
\draw[thick] (Ea) -- (Eb) -- (Ec) -- (Ed);
\draw[thick] (24,0) node {(E)};

\end{tikzpicture}
\end{center}
\caption{The five types of $P_4$'s}\label{fig:types} 
\end{figure}
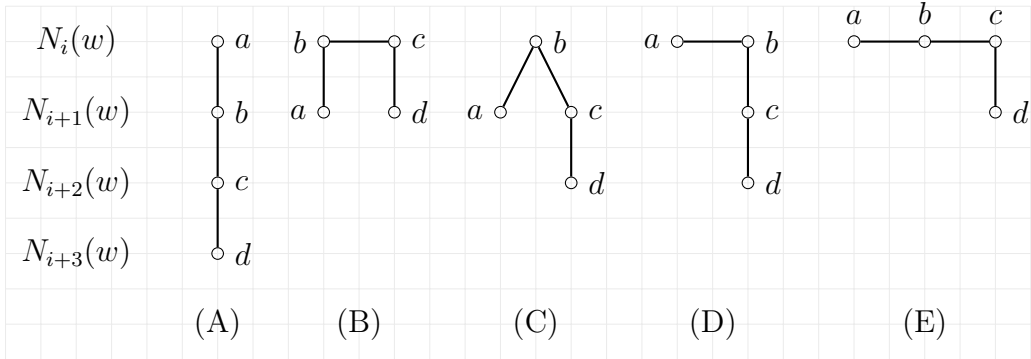

Thus, any induced $P_4\, abcd$ of $G$ is of exactly one of the following five types for some index $i\ge 0$ (up to re-labeling of the vertices; see also Figure~\ref{fig:types}):
\begin{itemize}
 \item[(A)] $a\in N_i(w),\, b\in N_{i+1}(w),\, c\in N_{i+2}(w)$, and $d\in N_{i+3}(w)$.
 \item[(B)] $b, c\in N_i(w)$, and $a,d\in N_{i+1}(w)$.
 \item[(C)] $b\in N_i(w),\, a,c\in N_{i+1}(w)$, and $d\in N_{i+2}(w)$.
 \item[(D)] $a,b\in N_i(w),\, c\in N_{i+1}(w)$, and $d\in N_{i+2}(w)$.
 \item[(E)] $a,b,c\in N_i(w)$, and $d\in N_{i+1}(w)$.
\end{itemize}

We now orient $G$ as follows. First, for each edge $uv$ with $u\in N_i(w)$ and $v\in N_{i+1}(w)$:
\begin{itemize}
 \item if $i = 0\!\mod 4$ or $i=1\!\mod 4$, orient the edge $uv$ \lq forward\rq, i.e., $u\to v$
   is the corresponding directed edge;
 \item otherwise, orient the edge $uv$ \lq backward\rq, i.e., $v\to u$ is the corresponding directed edge.
\end{itemize}

Clearly, in this partial orientation of $G$, any induced $P_4$ of type (A) or (B) is good, i.e., the orientation puts the two end-edges in opposition. 

Next, we orient the remaining edges as follows: Let $uv$ be an edge such that $u,v\in N_i(w)$ for some $i\ge 1$. If $uv$ is not an end-edge of any induced $P_4$, orient $uv$ arbitrarily. If $uv$ is an end-edge of an induced $P_4$, say $P$, then $P$ must be of type (D) or (E). In this case, $uv$ gets the direction that is in opposition to the direction of the other end-edge of $P$. 

In the proofs of Lemmas \ref{lem:h1oppo} and \ref{lem:final} below, the vertex $w$ will be chosen appropriately, and we will see that this orientation is well-defined, i.e., the other induced $P_4$s of types (D) and (E) are good, too. We will also see that any induced $P_4$ of type (C) is also good, provided $G$ is a $\mathsf{T}_k$-free $(k\ge 1)$ and $(\mathsf{G_1, G_2})$-free ptolemaic graph.

\begin{lemma}\label{lem:h1oppo}
Let $G$ be a $(\mathsf{G_1},H_1)$-free ptolemaic graph containing an induced $P_5$. Then $G$ is an opposition graph.
\end{lemma}
\proof
Let $w$ be the midpoint of an induced $P_5$, say $w_1w_2ww_4w_5$. Write $N_i$ for $N_i(w)$.
We first show that any $P_4\, uvxy$ of type (C) is good. Let $i\ge 0$ such that $v\in N_i,\, u,x\in N_{i+1}$, and $y\in N_{i+2}$. Then $i\le 1$, otherwise there would exist vertices $v'\in N_{i-1}\cap N(v), v''\in N_{i-2}\cap N(v')$ such that $v'', v', v, u, x$ and $y$ induce an $H_1$.

Assume $i=1$. Let $z\in\{w_2,w_4\}\setminus\{v\}$. If $v$ is non-adjacent to $z$, then $u, x$ are also non-adjacent to $z$ (as $G$ is chordal), hence $u,v,w,x,y$ and $z$ induce an $H_1$. 
Thus, $v$ is adjacent to $z$, implying $v\not= w_2,w_4$ and $v$ is adjacent to both $w_2, w_4$. Then, $u, x$ are non-adjacent to $w_2, w_4$ (as $G$ is gem-free chordal). This implies that $w_1\not=u, x$; moreover, since $G$ is chordal, $w_1$ is non-adjacent to $u,v,x,y$. Hence $w_1,w_2, u, v, x$ and $y$ induce an $H_1$.

So, $i=0$, and $v=w$. Thus we have $v\to u$ and $x\to y$, i.e., the $P_4\, uvxy$ is good.

Next, we show that the orientation for edges $uv$ with $u,v\in N_i$ for some $i\ge 1$ is well defined. Let $uv$ be an end-edge of an induced $P_4$, say $P\, uvxy$. If $P$ is the only $P_4$ containing $uv$, then the orientation of $uv$, implied by the one of $xy$, is well defined. Otherwise, we have two cases below. Let us assume that $xy$ is oriented by $x\to y$.

\smallskip\noindent
\textsc{Case 1:} $P$ is of type (E).
In this case, $u,v,x\in N_i$ and $y\in N_{i+1}$.\\
By Observation~\ref{obs}, there is a vertex $z\in N_{i-1}$ adjacent to $u,v$ and $x$. Thus, $uzxy$ is an induced $P_4$ of type (C) which has been shown to be good. In particular, $uz$ is oriented $z\to u$.
Now consider another induced $P_4$ containing $uv$, say $Q$. Assume $Q\, vuu'u''$ is of type (D).
Then $zuu'u''$ is of type (A), hence $u'u''$ is oriented $u''\to u'$. So $P$ and $Q$ imply the same orientation of $uv$.
Assume $Q\, uvv'v''$ is of type (D). Then $v'$ and $v''$ are not adjacent to $x, y$ (as $G$ is gem-free chordal), and $u, v, v', v'', x$ and $y$ induce an $H_1$. Thus such a $P_4$ of type (D) cannot exist. At this point, $Q$ is not of type (D), and we may assume that $Q$ is of type (E). If $Q$ is $vuu'u''$, then $u',u,v,x$ induce a $C_4$ (as no $P_4$ has all vertices in $N_i$) which is impossible as $G$ is chordal. If $Q$ is  $uvv'v''$, then $P$ and $Q$ imply the same orientation of $uv$. Case 1 is settled.

\smallskip\noindent
\textsc{Case 2:} $P$ is of type (D).
In this case, $u,v\in N_i, x\in N_{i+1}$ and $y\in N_{i+2}$.\\
By Observation~\ref{obs}, there is a vertex $z\in N_{i-1}$ adjacent to $u$ and $v$. Let $Q$ be another induced $P_4$ containing $uv$. By Case 1, we may assume that $Q$ is of type (D), and in order to create a conflict let $Q$ be $vuu'u''$. Then $1\le i\le 2$, otherwise there would exist vertices $z'\in N_{i-2}\cap N(z), z''\in N_{i-3}\cap N(z')$ such that $z'', z', z, P, Q$ induce a $\mathsf{G_1}$ (note that, as $G$ is chordal, there are no edges between $u', u''$ and $x, y$). We will see that the cases $i=1,2$ are also impossible.

Assume $i=1$. Then $z=w$. At least one of $w_2, w_4$ is different from $u, v$; let $w_2\not= u,v$, say. If $u$ or $v$ is non-adjacent to $w_2$, then $w_1,w_2,w,P, Q$ induce a $\mathsf{G_1}$ or there is a gem or $G$ is not chordal. So, $u$ and $v$ are adjacent to $w_2$, which implies that $w_4\not=u,v$; then, similarly, $u,v$ are adjacent to $w_4$. Now, $x$ is non-adjacent to $w_2$ and $w_4$: if $x$ is adjacent to $w_2$ and $w_4$, then $G[w,w_2,w_4,x]$ is a $C_4$; if $xw_2\in E(G)$ and $xw_4\not\in E(G)$, say, then $G[x, w_2, w, w_4, v]$ is a gem. Thus, $v,x,y,w_1,w_2$ and $w_4$ induce an $H_1$.

Assume $i=2$. Then $wz\in E(G)$, and let without loss of generality that $z\not= w_2$. Then, $z$ is adjacent to $w_2$: if $zw_2\not\in E(G)$, $w_1$ would be non-adjacent to $z,v,x,u,u'$ (otherwise $G$ would contain an induced $C_k, k\ge 4$), and thus $G$ would contain an induced $\mathsf{G_1}$. Thus, $z\not=w_4$, and by symmetry, $z$ is adjacent to $w_4$. Now, $u$ is non-adjacent to $w_2$ and $w_4$: if $u$ is adjacent to both $w_2,w_4$, then $G[w,w_2,w_4,u]$ is a $C_4$; if $uw_2\in E(G)$ and $uw_4\not\in E(G)$, say, then $G[u,w_2,w,w_4,z]$ is a gem. By symmetry, $v$ is non-adjacent to $w_2,w_4$. Moreover, $w_1$ is non-adjacent to $z$ (otherwise $G[w_1,w_2,w,w_4,z]$ would be a gem). Thus, $w_1, w_2, z, P, Q$ induce a $\mathsf{G_1}$ (note that, as $G$ is chordal, $w_1$ is non-adjacent to $u, v, u', x$).

Case 2 is settled, and the proof of Lemma~\ref{lem:h1oppo} is complete.\qed

\begin{lemma}\label{lem:final}
Let $G$ be a $\mathsf{T}_k$-free $(k\ge 1)$ and $(\mathsf{G_1, G_2})$-free ptolemaic graph containing an induced $H_1$. Then $G$ is an opposition graph.
\end{lemma}
\proof
The main difficulty in this case is the choice of the root vertex $w$ in defining the
sets $N_i(w)$ as in the proof of Lemma~\ref{lem:h1oppo}.

\begin{figure}[ht] 
\begin{center}
\begin{tikzpicture}[scale=.4] 
\tikzstyle{vertexvk}=[draw,circle,inner sep=1.55pt,fill=gray!30];
\tikzstyle{vertex}=[draw,circle,inner sep=1.55pt] 
\node[vertex] (v0)   at (0,2) [label=left:\footnotesize $v_0$] {};
\node[vertex] (v0')  at (2,2) [label=below:\footnotesize $\hspace*{6mm}v_0'$] {};
\node[vertex] (v0'') at (2,0) [label=below:\footnotesize $\hspace*{6mm}v_0''$] {};
\node[vertexvk] (v1)   at (4,2) [label=below:\footnotesize $v_1$] {};
\node[vertex] (v1')  at (6,2) [label=below:\footnotesize $\hspace*{6mm}v_1'$] {};
\node[vertex] (v1'') at (6,0) [label=below:\footnotesize $\hspace*{6mm}v_1''$] {};
\node[vertexvk] (v2)   at (4,4.7) [label=above:\footnotesize $v_2$] {};
\node[vertex] (v2')  at (8,2) [label=below:\footnotesize $\hspace*{6mm}v_2'$] {};
\node[vertex] (v2'') at (8,0) [label=below:\footnotesize $\hspace*{6mm}v_2''$] {};

\draw (v2) -- (v0); \draw (v2) -- (v0'); \draw (v2) -- (v1);
\draw (v2) -- (v1'); \draw (v2) -- (v2');

\draw (v0'') -- (v0') -- (v1) -- (v1') -- (v1'');
\draw (v2') -- (v2'');

\draw (v0) to[bend angle=20, bend left] (v1);

\draw (6,4.3) node {\footnotesize $H_2$};
\end{tikzpicture}
\qquad
\begin{tikzpicture}[scale=.4] 
\tikzstyle{vertexvk}=[draw,circle,inner sep=1.55pt,fill=gray!30];
\tikzstyle{vertex}=[draw,circle,inner sep=1.55pt] 
\node[vertex] (v0)   at (2,0) [label=below:\footnotesize $\hspace*{6mm}v_0$] {};
\node[vertex] (v0')  at (0,2) [label=below:\footnotesize $\hspace*{6mm}v_0'$] {};
\node[vertex] (v0'') at (0,0) [label=below:\footnotesize $\hspace*{6mm}v_0''$] {};
\node[vertexvk] (v1)   at (2,2) [label=below:\footnotesize $\hspace*{6mm}v_1$] {};
\node[vertex] (v1')  at (4,2) [label=below:\footnotesize $\hspace*{6mm}v_1'$] {};
\node[vertex] (v1'') at (4,0) [label=below:\footnotesize $\hspace*{6mm}v_1''$] {};
\node[vertexvk] (v2)   at (3,4.7) [label=above:\footnotesize $v_2$] {};
\node[vertex] (v2')  at (6,2) [label=below:\footnotesize $\hspace*{6mm}v_2'$] {};
\node[vertex] (v2'') at (6,0) [label=below:\footnotesize $\hspace*{6mm}v_2''$] {};

\draw (v2) -- (v0'); \draw (v2) -- (v1);
\draw (v2) -- (v1'); \draw (v2) -- (v2');

\draw (v0'') -- (v0') -- (v1) -- (v1') -- (v1'');
\draw (v0) -- (v1);
\draw (v2') -- (v2'');

\draw (5,4.3) node {\footnotesize $H_2^-$};
\end{tikzpicture}
\end{center}
\begin{center}
\begin{tikzpicture}[scale=.4]
\tikzstyle{vertexvk}=[draw,circle,inner sep=1.55pt,fill=gray!30];
\tikzstyle{vertex}=[draw,circle,inner sep=1.55pt] 
\node[vertex] (v0)   at (0,2) [label=left:\footnotesize $v_0$] {};
\node[vertex] (v0')  at (2,2) [label=below:\footnotesize $\hspace*{6mm}v_0'$] {};
\node[vertex] (v0'') at (2,0) [label=below:\footnotesize $\hspace*{6mm}v_0''$] {};
\node[vertexvk] (v1)   at (4,2) [label=below:\footnotesize $v_1$] {};
\node[vertex] (v1')  at (6,2) [label=below:\footnotesize $\hspace*{6mm}v_1'$] {};
\node[vertex] (v1'') at (6,0) [label=below:\footnotesize $\hspace*{6mm}v_1''$] {};
\node[vertexvk] (v2)   at (8,2) [label=below:\footnotesize $v_2$] {};
\node[vertex] (v2')  at (10,2) [label=below:\footnotesize $\hspace*{6mm}v_2'$] {};
\node[vertex] (v2'') at (10,0) [label=below:\footnotesize $\hspace*{6mm}v_2''$] {};
\node[vertexvk] (v3)   at (6,5.6) [label=above:\footnotesize $v_3$] {};
\node[vertex] (v3')  at (12,2) [label=below:\footnotesize $\hspace*{6mm}v_3'$] {};
\node[vertex] (v3'') at (12,0) [label=below:\footnotesize $\hspace*{6mm}v_3''$] {};

\draw (v3) -- (v0); \draw (v3) -- (v0'); \draw (v3) -- (v1);
\draw (v3) -- (v1'); \draw (v3) -- (v2); \draw (v3) -- (v2');
\draw (v3) -- (v3') -- (v3''); 

\draw (v0'') -- (v0') -- (v1) -- (v1') -- (v2) -- (v2') -- (v2'');
\draw (v1') -- (v1'');

\draw (v0) to[bend angle=20, bend left] (v1);
\draw (v0) to[bend angle=25, bend left] (v2);
\draw (v0') to[bend angle=22.5, bend left] (v2);
\draw (v1) to[bend angle=20, bend left] (v2);

\draw (10,5) node {\footnotesize $H_3$};
\end{tikzpicture}
\qquad
\begin{tikzpicture}[scale=.4]
\tikzstyle{vertexvk}=[draw,circle,inner sep=1.55pt,fill=gray!30];
\tikzstyle{vertex}=[draw,circle,inner sep=1.55pt] 
\node[vertex] (v0)   at (2,0) [label=below:\footnotesize $\hspace*{6mm}v_0$] {};
\node[vertex] (v0')  at (0,2) [label=below:\footnotesize $\hspace*{6mm}v_0'$] {};
\node[vertex] (v0'') at (0,0) [label=below:\footnotesize $\hspace*{6mm}v_0''$] {};
\node[vertexvk] (v1)   at (2,2) [label=below:\footnotesize $\hspace*{6mm}v_1$] {};
\node[vertex] (v1')  at (4,2) [label=below:\footnotesize $\hspace*{6mm}v_1'$] {};
\node[vertex] (v1'') at (4,0) [label=below:\footnotesize $\hspace*{6mm}v_1''$] {};
\node[vertexvk] (v2)   at (6,2) [label=below:\footnotesize $v_2$] {};
\node[vertex] (v2')  at (8,2) [label=below:\footnotesize $\hspace*{6mm}v_2'$] {};
\node[vertex] (v2'') at (8,0) [label=below:\footnotesize $\hspace*{6mm}v_2''$] {};
\node[vertexvk] (v3)   at (5,5.6) [label=above:\footnotesize $v_3$] {};
\node[vertex] (v3')  at (10,2) [label=below:\footnotesize $\hspace*{6mm}v_3'$] {};
\node[vertex] (v3'') at (10,0) [label=below:\footnotesize $\hspace*{6mm}v_3''$] {};

\draw (v3) -- (v0'); \draw (v3) -- (v1);
\draw (v3) -- (v1'); \draw (v3) -- (v2); \draw (v3) -- (v2');
\draw (v3) -- (v3') -- (v3''); 

\draw (v0'') -- (v0') -- (v1) -- (v1') -- (v2) -- (v2') -- (v2'');
\draw (v0) -- (v1);
\draw (v1') -- (v1'');

\draw (v0') to[bend angle=22.5, bend left] (v2);
\draw (v1) to[bend angle=20, bend left] (v2);

\draw (8,5) node {\footnotesize $H_3^-$};
\end{tikzpicture}
\end{center}

\begin{center}
\begin{tikzpicture}[scale=.35]
\tikzstyle{vertexvk}=[draw,circle,inner sep=1.55pt,fill=gray!30];
\tikzstyle{vertex}=[draw,circle,inner sep=1.55pt] 
\node[vertex] (v0)   at (0,2) [label=left:{\footnotesize $v_0$}] {};
\node[vertex] (v0')  at (2,2) [label=below:{\footnotesize $\hspace*{5mm}v_0'$}] {};
\node[vertex] (v0'') at (2,0) [label=below:{\footnotesize $\hspace*{5mm}v_0''$}] {};
\node[vertexvk] (v1)   at (4,2) [label=below:{\footnotesize $v_1$}] {};
\node[vertex] (v1')  at (6,2) [label=below:{\footnotesize $\hspace*{5mm}v_1'$}] {};
\node[vertex] (v1'') at (6,0) [label=below:{\footnotesize $\hspace*{5mm}v_1''$}] {};
\node[vertexvk] (v2)   at (8,2) [label=below:{\footnotesize $v_2$}] {};
\node[vertex] (v2')  at (10,2) [label=below:{\footnotesize $\hspace*{5mm}v_2'$}] {};
\node[vertex] (v2'') at (10,0) [label=below:{\footnotesize $\hspace*{5mm}v_2''$}] {};
\node[vertexvk] (v3)   at (12,2) [label=below:{\footnotesize $v_3$}] {};
\node[vertex] (v3')  at (14,2) [label=below:{\footnotesize $\hspace*{5mm}v_3'$}] {};
\node[vertex] (v3'') at (14,0) [label=below:{\footnotesize $\hspace*{5mm}v_3''$}] {};
\node[vertexvk] (v4)   at (8,7.5) [label=above:{\footnotesize $v_4$}] {};
\node[vertex] (v4')  at (16,2) [label=below:{\footnotesize $\hspace*{5mm}v_4'$}] {};
\node[vertex] (v4'') at (16,0) [label=below:{\footnotesize $\hspace*{5mm}v_4''$}] {};

\draw (v4) -- (v0); \draw (v4) -- (v0'); \draw (v4) -- (v1);
\draw (v4) -- (v1'); \draw (v4) -- (v2); \draw (v4) -- (v2');
\draw (v4) -- (v3); \draw (v4) -- (v3'); \draw (v4) -- (v4') -- (v4''); 

\draw (v0'') -- (v0') -- (v1) -- (v1') -- (v2) -- (v2') -- (v3) -- (v3') -- (v3'');
\draw (v1') -- (v1''); \draw (v2') -- (v2'');

\draw (v0) to[bend angle=20, bend left] (v1);
\draw (v0) to[bend angle=25, bend left] (v2);
\draw (v0) to[bend left] (v3);
\draw (v0') to[bend angle=22.5, bend left] (v2);
\draw (v0') to[bend angle=27.5, bend left] (v3);
\draw (v1) to[bend angle=20, bend left] (v2);
\draw (v1) to[bend angle=25, bend left] (v3);
\draw (v1') to[bend angle=22.5, bend left] (v3);
\draw (v2) to[bend angle=20, bend left] (v3);

\draw (13,7) node {\footnotesize $H_4$};
\end{tikzpicture}
\quad
\begin{tikzpicture}[scale=.35]
\tikzstyle{vertexvk}=[draw,circle,inner sep=1.55pt,fill=gray!30];
\tikzstyle{vertex}=[draw,circle,inner sep=1.55pt] 
\node[vertex] (v0)   at (2,0) [label=below:{\footnotesize $\hspace*{5mm}v_0$}] {};
\node[vertex] (v0')  at (0,2) [label=below:{\footnotesize $\hspace*{5mm}v_0'$}] {};
\node[vertex] (v0'') at (0,0) [label=below:{\footnotesize $\hspace*{5mm}v_0''$}] {};
\node[vertexvk] (v1)   at (2,2) [label=below:{\footnotesize $\hspace*{5mm}v_1$}] {};
\node[vertex] (v1')  at (4,2) [label=below:{\footnotesize $\hspace*{5mm}v_1'$}] {};
\node[vertex] (v1'') at (4,0) [label=below:{\footnotesize $\hspace*{5mm}v_1''$}] {};
\node[vertexvk] (v2)   at (6,2) [label=below:{\footnotesize $v_2$}] {};
\node[vertex] (v2')  at (8,2) [label=below:{\footnotesize $\hspace*{5mm}v_2'$}] {};
\node[vertex] (v2'') at (8,0) [label=below:{\footnotesize $\hspace*{5mm}v_2''$}] {};
\node[vertexvk] (v3)   at (10,2) [label=below:{\footnotesize $v_3$}] {};
\node[vertex] (v3')  at (12,2) [label=below:{\footnotesize $\hspace*{5mm}v_3'$}] {};
\node[vertex] (v3'') at (12,0) [label=below:{\footnotesize $\hspace*{5mm}v_3''$}] {};
\node[vertexvk] (v4)   at (7,6.5) [label=above:{\footnotesize $v_4$}] {};
\node[vertex] (v4')  at (14,2) [label=below:{\footnotesize $\hspace*{5mm}v_4'$}] {};
\node[vertex] (v4'') at (14,0) [label=below:{\footnotesize $\hspace*{5mm}v_4''$}] {};

\draw (v4) -- (v0'); \draw (v4) -- (v1);
\draw (v4) -- (v1'); \draw (v4) -- (v2); \draw (v4) -- (v2');
\draw (v4) -- (v3); \draw (v4) -- (v3'); \draw (v4) -- (v4') -- (v4''); 

\draw (v0'') -- (v0') -- (v1) -- (v1') -- (v2) -- (v2') -- (v3) -- (v3') -- (v3'');
\draw (v1') -- (v1''); \draw (v2') -- (v2'');
\draw (v1) -- (v0);

\draw (v0') to[bend angle=22.5, bend left] (v2);
\draw (v0') to[bend angle=27.5, bend left] (v3);
\draw (v1) to[bend angle=20, bend left] (v2);
\draw (v1) to[bend angle=25, bend left] (v3);
\draw (v1') to[bend angle=22.5, bend left] (v3);
\draw (v2) to[bend angle=20, bend left] (v3);

\draw (11,6) node {\footnotesize $H_4^-$};
\end{tikzpicture}
\end{center}
\caption{The graphs $H_k$ and $H_k^-$}\label{fig:Hk} 
\end{figure}
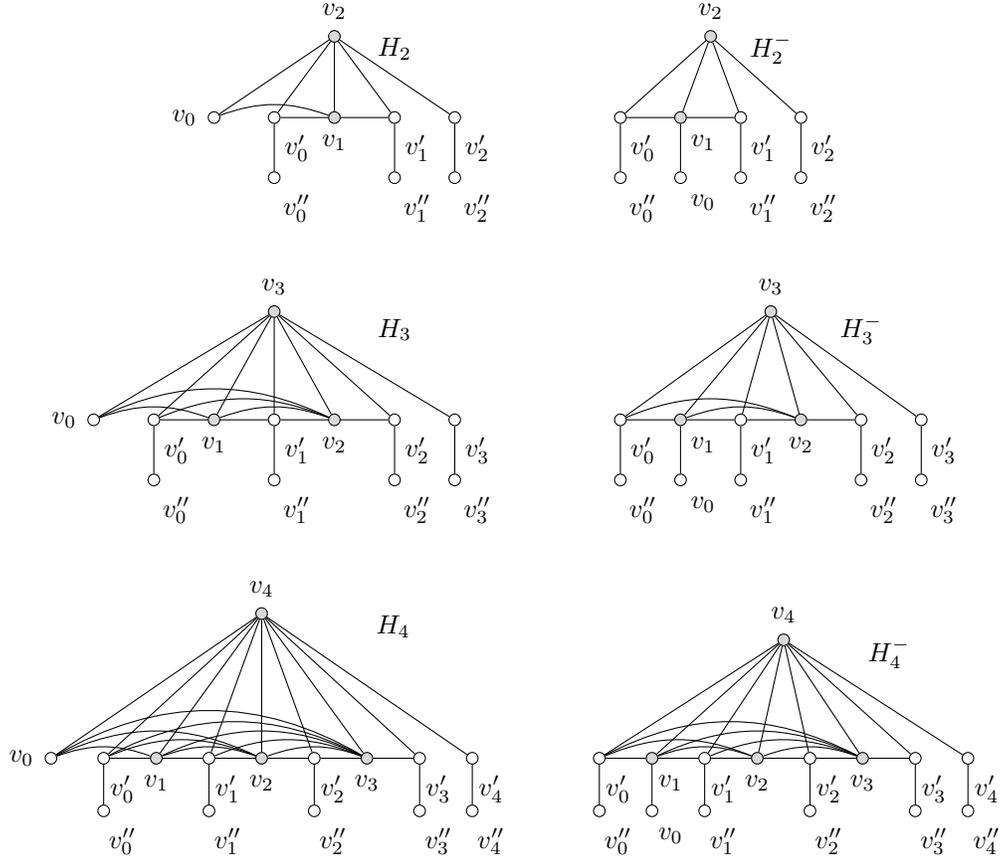

First, let $H_1^-=H_1$ and define the graphs $H_{k+1}$ and $H_{k+1}^-$, $k\ge 1$, as follows. $H_{k+1}$ (respectively, $H_{k+1}^-$) is obtained from $H_k$ (respectively, $H_k^-$) by taking three new vertices $v_{k+1}, v_{k+1}', v_{k+1}''$, and joining $v_{k+1}$ to all vertices in $N[v_k]$ (respectively, in $N[v_k]\setminus\{v_0\}$), $v_{k+1}'$ to $v_{k+1}$, and $v_{k+1}''$ to $v_{k+1}'$. Note that, for $k\ge 2$, $H_k^-$ is obtained from $H_k$ by deleting all edges $v_iv_0, 2\le i\le k$. See Figure~\ref{fig:Hk}.

Note also that, for all $k\ge 1$, each of $H_{k}$ and $H_k^-$ is an opposition graph admitting exactly two opposition orientations: In the first one, $v_k$ is a source. The second one is the reverse of the other (in which, $v_k$ is a sink). In any opposition orientation of $H_{k}$ and of $H_k^-$, each $v_i$, $1\le i\le k-1$, is neither a source nor a sink.

Now, let $k\ge 1$ be the largest integer such that $G$ contains an induced $H=H_k$ or an induced $H=H_k^-$. Then let $w$ be the vertex $v_{k}$ of such an $H$. Again, write $N_i$ for $N_i(w)$. We first show:

\begin{observation}\label{obs2}
Let $w_1$ be a vertex in $N_1$ such that there exists a vertex $w_2\in N(w_1)\cap N_2$ and a vertex $w_3\in N(w_2)\cap N_3$. Then $w_1$ is non-adjacent to a vertex $v_t'\in V(H)$ for some $0\le t\le k$.
\end{observation}
{\em Proof of Observation~\ref{obs2}:\/} 
Note that $v_k=w$ is the only vertex in $H$ adjacent to all $v_0', \ldots, v_k'$. Hence the statement of the observation is trivially true in case $w_1\in V(H)$.

So, let $w_1\not\in V(H)$, and suppose the contrary that $w_1$ is adjacent to all $v_0', \ldots, v_k'$. Then 
\begin{equation*}
\text{$w_1$ is adjacent to all $v_1, \ldots, v_k$,} 
\end{equation*}
for, if $w_1$ was non-adjacent to a vertex $v_s$ for some $1\le s\le k$, then $w_1, v_0', v_s, v_s'$ would induce a $C_4$, and 
\begin{equation*}
\text{$w_1$ is non-adjacent to all $v_0'', \ldots, v_k''$,} 
\end{equation*}   
for, if $w_1$ was adjacent to a vertex $v_s''$ for some $0\le s\le k$, then $w_1, v_s'', v_s', v_k, v_k'$ (if $s=0$) or $w_1, v_s'', v_s', v_k, v_0'$ (if $s\not=0$) would induce a gem. Then, $w_2\not\in H$ since $N(w_1)\cap H\subseteq N(w)$. Moreover,
\begin{equation*}
\text{$w_2$ is non-adjacent to all $v_1, \ldots, v_k$,} 
\end{equation*}
for, if $w_2$ was adjacent to a vertex $v_s$ for some $1\le s\le k-1$, then $w_2$ would be non-adjacent to $v_k'$ (otherwise $G[w_2, v_s, v_k, v_k']$ would be a $C_4$) and $G[w_1, w_2, v_s, v_k, v_k']$ would be a gem, and 
\begin{equation*}
\text{$w_2$ is non-adjacent to all $v_0', \ldots, v_k'$,} 
\end{equation*}
for, if $w_2$ was adjacent to $v_s'$ for some $1\le s\le k$, then $w_2$ would be non-adjacent to $v_{s-1}'$ (otherwise $G[w_2, v_s', v_k, v_{s-1}']$ would be a $C_4$) and $G[w_1, w_2, v_s'$, $v_k, v_{s-1}']$ would be a gem. It follows that $w_2$ is non-adjacent to $v_0'$, otherwise $G[w_1, w_2, v_0', v_k, v_k']$ would be a gem. Furthermore    
\begin{equation*}
\text{$w_2$ is non-adjacent to all $v_0'', \ldots, v_k''$,} 
\end{equation*}
for, if $w_2$ was adjacent to $v_s''$ for some $0\le s\le k$, then $w_2, v_s'', v_s'$ and $w_1$ would induce a $C_4$. 

Similarly, we can show that $w_3$ is non-adjacent to all $v_0'', \ldots, v_k''$, and that, in case $H=H_k^-$, $w_3$ is also non-adjacent to $v_0$. It follows that
\begin{equation*}
\text{$w_3$ is non-adjacent to all vertices of $H$.} 
\end{equation*}
We now distinguish two cases. 

Suppose $k=1$. In this case, $H$ and $v_2:= w_1, v_2':= w_2, v_2'':= w_3$ induce an $H_2$ (if $w_1$ is adjacent to $v_0$) or an $H_2^-$ (otherwise), contradicting the choice of $k=1$.

Suppose $k\ge 2$. In this case, if $H=H_k$, then $w_1$ must be adjacent to $v_0$, otherwise $v_0, v_1, w_1, v_k'$ and $v_k$ would induce a gem. But then $H$ and $v_{k+1}:= w_1, v_{k+1}':= w_2, v_{k+1}'':= w_3$ would induce an $H_{k+1}$, contradicting the choice of $k$. If $H=H_k^-$, then $w_1$ must be non-adjacent to $v_0$, otherwise $v_0, v_1, v_k, v_k'$ and $w_1$ would induce a gem. But then $H$ and $v_{k+1}:= w_1, v_{k+1}':= w_2, v_{k+1}'':= w_3$ would induce an $H_{k+1}^-$, a contradiction again.

Thus, $w_1$ cannot be adjacent to all $v_0', \ldots, v_k'$, and Observation~\ref{obs2} is proved.

\smallskip
We next show that any induced $P_4$ of type (C) is good. This will be implied by the definition of the orientation and the following fact.
\begin{observation}\label{typeC} 
Let $P\, uvxy$ be an induced $P_4$ of type (C) and $i\ge 0$ such that $v\in N_i,\, u,x\in N_{i+1}$, and $y\in N_{i+2}$. Then $i$ is even. 
\end{observation}
{\em Proof of Observation~\ref{typeC}:\/} 
Suppose, for a contradiction, that $i$ is odd. Let $Q\, w_0w_1\ldots w_i$ be a chordless path of odd length $i$, connecting $w_0=w=v_k$ and $w_i=v$ (thus, $w_j\in N_j$ for all $0\le j\le i$). Set $w_{i+1}=x, w_{i+2}=y$.

\smallskip
Consider first the case $w_1\in H$.  
Suppose $w_1=v_s$ for some $0\le s\le k-1$. Then, since $w_1w_2\in E(G)$ and $v_sv_j''\not\in E(G)$ for all $0\le j\le k$, $w_2\not\in H$. Moreover, there are no edges between $w_2, w_3$ and $v_k', v_k''$ (otherwise $G[w_3,w_2,v_s, v_k, v_k',v_k'']$ would contain a $C_4, C_5$, or a $C_6$). 
Now, if $s=0$, then $P,Q,v_0',v_k',v_k''$ induce a $\mathsf{T}_{i+1}$. If $s\not=0$, then there are no edges between $w_2,w_3$ and $v_{s-1}', v_s',v_s''$ (otherwise $G[w_3,w_2,v_{s-1}', v_s, v_s',v_s'',v_k]$ would contain a gem, a $C_4$ or a $C_5$), hence $w_3,w_2,v_{s-1}', v_s, v_s',v_s'',v_k, v_k',v_k''$ induce a $\mathsf{G_2}$. In any case we have a contradiction.

Suppose $w_1=v_s'$ for some $0\le s\le k$. Then, $w_2\not\in H-v_s''$ (as $w_1w_2\in E(G)$ and $v_s'v_j''\not\in E(G)$ for all $j\not=s$), and there are no edges between $w_2,w_3$ and $v_t',v_t''$ for any $t\not=s$ (otherwise $G[w_3,w_2,v_{s}', v_k, v_t',v_t'']$ would contain a $C_4, C_5$ or a $C_6$). Thus, if $k\ge 2$, then $P,Q,v_{t_1}', v_{t_1}'',v_{t_2}'$ induce a $\mathsf{T}_{i+1}$ for any pair $t_1\not=t_2\in\{0,\ldots, k\}\setminus\{s\}$. In case $k=1$, $P,Q,v_0,v_t', v_t''$ induce a $\mathsf{T}_{i+1}$ ($w_2$ and $w_3$ are non-adjacent to $v_0$ because $G$ is chordal). Again, in any case, we have a contradiction.

\smallskip
So, we may assume that $w_1\not\in H$. 
Let $T'$ be the set of all vertices in $\{v_0',\ldots, v_k'\}$ such that $w_1$ is non-adjacent to all vertices in $T'$. 
By Observation~\ref{obs2}, $T'\not=\emptyset$. Set $T''=\{v_i'' \mid v_i\in T'\}$. Note that $w_2\not\in T''$ (if $w_2=v_t''$ for some $t$, then $G[v_t'',v_t',v_k,w_1]$ is a $C_4$), and there are no edges between $w_2,w_3$ and $T'\cup T''$ (otherwise $G[\{w_3,w_2,w_1,v_k\}\cup T'\cup T'']$ would contain a $C_4, C_5$ or a $C_6$). Thus, if $|T'|\ge 2$, then $P,Q,v_t',v_t'',v_s'$ induce a $\mathsf{T}_{i+1}$ for any $t\not=s$ with $v_t', v_s'\in T'$, a contradiction.

So, let $T'=\{v_t'\}$ for some $0\le t\le k$. Suppose $k\ge 2$, and let $s_1\not= s_2\in\{0,\ldots,k\}\setminus\{t\}$. Then, since $w_1$ is adjacent to $v_i'$ for all $i\not= t$, $w_2$ is non-adjacent to $v_{s_1}'$ and $v_{s_2}'$ (otherwise $G[w_1,w_2, v_{s_1}',v_k,v_{s_2}']$ would contain a $C_4$ or a gem), implying $w_2\not\in\{v_{s_1}'',v_{s_2}''\}$. Furthermore, $w_2,w_3$ are non-adjacent to $v_{s_1}''$ (otherwise $G[w_3,w_2,w_1,v_{s_1}',v_{s_2}']$ would contain a $C_4$ or $C_5$). Thus, $w_3,w_2,w_1,v_k,v_t',v_t'',v_{s_1}',v_{s_1}'',v_{s_2}'$ induce a $\mathsf{G_2}$, a contradiction.

So, let $k=1$, and assume without loss of generality that $t=0$. Now, if $w_1$ is non-adjacent to $v_0$, then $w_2,w_3$ are non-adjacent to $v_0$ (otherwise $G[w_3,w_2,w_1,v_1,v_0]$ would contain a $C_4$ or $C_5$), and $P,Q,v_0,v_o',v_0''$ induce a $\mathsf{T}_{i+1}$. If $w_1$ is adjacent to $v_0$, then $w_2$ is non-adjacent to $v_0,v_1'$ (otherwise $G[w_2,w_1,v_1,v_1',v_0]$ would contain a $C_4$ or a gem), hence $w_2\not= v_1''$, and $w_2$ and $w_3$ are non-adjacent to $v_1''$ (otherwise $G[w_3,w_2,w_1,v_1'.v_1'']$ would contain a $C_4$ or $C_5$). Thus, $H=H_1$ together with $w_1, w_2, w_3$ induce a $\mathsf{G_2}$. In any case, we have a contradiction.

Thus, $i$ must be even, and the proof of Observation~\ref{typeC} is complete.

\medskip
It follows from the definition of the orientation and Observation~\ref{typeC}  that all induced $P_4$s of type (C) are good. 

We finally show that the orientation for edges $uv$ with $u,v\in N_i$ for some $i\ge 1$ is well defined (that is, all other induced $P_4$s of types (D) and (E) are good). Let $uv$ be an end-edge of an induced $P_4$, say $P\, uvxy$. If $P$ is the only $P_4$ containing $uv$, then the orientation of $uv$, implied by the one of $xy$, is well defined. Otherwise, we have two cases below. Let us assume that $xy$ is oriented by $x\to y$.

\smallskip\noindent
\textsc{Case 1:} $P$ is of type (E).
In this case, $u,v,x\in N_i$ and $y\in N_{i+1}$.\\
By Observation~\ref{obs}, there is a vertex $z\in N_{i-1}$ adjacent to $u,v$ and $x$. Thus, $uzxy$ is an induced $P_4$ of type (C) which has been shown to be good. In particular, $uz$ is oriented $z\to u$.
Now consider another induced $P_4$ containing $uv$, say $Q$. Assume $Q\, vuu'u''$ is of type (D).
Then $zuu'u''$ is of type (A), hence $u'u''$ is oriented $u''\to u'$. So $P$ and $Q$ imply the same orientation of $uv$.

Assume $Q\, uvv'v''$ is of type (D). Note that there are no edges between $\{v', v''\}$ and $\{x,y\}$ (as $G$ is gem-free chordal). Hence $i-1\le 1$, otherwise $u,v,v',v'',x,y,z$, a vertex $z'\in N(z)\cap N_{i-2}$ and $z''\in N(z')\cap N_{i-3}$ would together induce a $\mathsf{G_2}$. Moreover, Observation~\ref{typeC} (for the $P_4\, uzxy$) implies that $i=1$. Thus, $z=w=v_k$. By Observation~\ref{obs2}, $v$ is non-adjacent to a vertex $v'_t\in V(H)\setminus\{v\}$ for some $0\le t\le k$.  Then $v'_t\not\in\{u,x\}$ and $G[u,v,x,y,v',v'',v_k,v_t',v_t'']$ is a $\mathsf{G_2}$ or contains an induced gem or an induced cycle of length at least $4$, a contradiction. 

So, $Q$ is not of type (D), and we may assume that $Q$ is of type (E). If $Q$ is $vuu'u''$, then $u',u,v,x$ induce a $C_4$ (as no $P_4$ has all vertices in $N_i$) which is impossible as $G$ is chordal. If $Q$ is  $uvv'v''$, then $P$ and $Q$ imply the same orientation of $uv$. Case 1 is settled.

\smallskip\noindent
\textsc{Case 2:} $P$ is of type (D).
In this case, $u,v\in N_i, x\in N_{i+1}$ and $y\in N_{i+2}$.\\
By Observation~\ref{obs}, there is a vertex $z\in N_{i-1}$ adjacent to $u$ and $v$. Let $Q$ be another induced $P_4$ containing $uv$. By Case 1, we may assume that $Q$ is of type (D). Let $Q$ be $vuu'u''$. Then, as $G$ is chordal, $y\not=u''$ and there are no edges between $u', u''$ and $x, y$. Moreover, $1\le i\le 2$, otherwise there would exist vertices $z'\in N_{i-2}\cap N(z), z''\in N_{i-3}\cap N(z')$ such that $z'', z', z, P, Q$ induce a $\mathsf{G_1}$. We will see that the cases $i=1,2$ are also impossible.

Assume $i=1$. Then $z=w=v_k$. By Observation~\ref{obs2}, $u$ is non-adjacent to some $v_{t}'$ and $v$ is non-adjacent to some $v_{s}'$ in $H$. If $t=s$, then $G[u,v,x,y,u',u'', v_k, v_t', v_t'']$ is a $\mathsf{G_1}$ (note that, as $G$ is chordal, $v_t''\not=u',x$ and there are no edges between $v_t', v_t''$ and $u,u',u'',v,x,y$). So, we may assume that $t\not= s$ and $u$ is adjacent to $v_{s}'$, $v$ is adjacent to $v_{t}'$. But then $v_k, v_t', v, u, v_s'$ induce a gem, a contradiction.

Assume $i=2$. Then $v_kz\in E(G)$, and by Observation~\ref{obs2}, $z$ is non-adjacent to a vertex $v_t'$ in $H$. As $G$ is chordal, $v_t'$ is non-adjacent to $u,v$. But then $G[u,v,x,y,u',u'',z, v_k, v_t']$ is a $\mathsf{G_1}$, a contradiction.

Case 2 is settled, and the proof of Lemma~\ref{lem:final} is completed.\qed

\begin{corollary}[\cite{Le2013}]
A tree is an opposition graph if and only if it is $\mathsf{T}_k$-free, $k\ge 1$.
\end{corollary}

\section{Coalition graphs}\label{sec:dhcoali}

If we define \lq generalized coalition graphs\rq\ by omitting the condition \lq acyclic\rq\ in the Definition~\ref{def:coali}, we unfortunately lose the perfectness of the graphs we obtain, as any cycle admits a \lq generalized coalition orientation\rq. Nevertheless, we do define a similar auxiliary graph, and it turns out that this is useful when discussing certain {\it hole-free\/} coalition graphs.

\begin{figure}[H]
\begin{center}
\begin{tikzpicture}[scale=.6] 
\tikzstyle{vertex}=[draw,circle,inner sep=1.5pt]
\node[vertex] (1) at (0,1)  {};
\node[vertex] (2) at (1,1)  {};
\node[vertex] (3) at (2,1)  {};
\node[vertex] (4) at (3,1)  {};
\node[vertex] (5) at (1.5,2)  {};
\node[vertex] (6) at (1.5,3)  {};

\draw (1) -- (2) -- (3) -- (4);
\draw (2) -- (5) -- (3);
\draw (5) -- (6);

\draw(1.5,.3) node {$\mathsf{N}$};
\end{tikzpicture}
\end{center}
\caption{The forbidden graph for distance-hereditary coalition graphs.}\label{fig:N}
\end{figure}
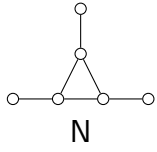

Given a graph $G=(V,E)$, the graph $\mathscr{C}(G)$ has
\begin{itemize}
 \item $\left\{(x,y)\mid \{x,y\} \text{ is an end-edge of an induced $P_4$ in $G$}\right\}$ as vertex set,
 \item two vertices $(x,y)$ and $(u,v)$ are adjacent in $\mathscr{C}(G)$ if $(x,y)=(v,u)$, or else
    $xyvu$\, or\, $vuxy$\, is an induced $P_4$ in $G$.
\end{itemize}
As an example, if $G$ is the graph $\mathsf{N}$ depicted in Figure~\ref{fig:N}, then $\mathscr{C}(G)$
is the complement of $C_6$ (the graph on the left side in Figure~\ref{fig:genoppo}).

Now it can be shown (cf. Theorem~\ref{thm:genoppo}) that $G$ is a \lq generalized coalition graph\rq\ if and only if $\mathscr{C}(G)$ is bipartite.
Moreover, we have the following theorem, similar to Theorem~\ref{thm:gemhouse} about $(\text{gem, house})$-free opposition graphs. 

\begin{theorem}\label{thm:gemhouseholecoali}
Let $G$ be $(\text{gem, house, hole})$-free. Then $G$ is a coalition graph if and only if $\mathscr{C}(G)$
is bipartite.
\end{theorem}
\proof For the non-trivial direction, let $\mathscr{C}(G)$ be bipartite and consider a bipartition $A, B$ of  $\mathscr{C}(G)$. Following the proof of Theorem~\ref{thm:gemhouse}, 
let $D(A)$ be the (partial) orientation of $G$ such that an edge $\{x,y\}$ of $G$ that is an end-edge of an induced $P_4$ is oriented $x\to y$ in $D(A)$ if $(x,y)\in A$, otherwise
$y\to x\in D(A)$. Recall that any induced $P_4\, xyuv$ in $G$ is \lq good\rq: it has $x\to y$ and $u\to v$ in $D(A)$ or else $y\to x$ and $v\to u$ in $D(A)$.

As in the proof of Theorem~\ref{thm:gemhouse}, we will show that $D(A)$ is acyclic. Assume, by contradiction, that $C\, v_1v_2 \ldots v_qv_1$ is a directed cycle in $D(A)$.
Note that, since $v_i\to v_{i+1}\in D(A)$, by definition of $D(A)$, $(v_i, v_{i+1})\in A$ for
all $i$. Choose such a cycle $C$ with minimum length $q\ge 3$. Then,
\begin{equation}\label{eq:coalition}
\text{no chord of $C$ is an end-edge of an induced $P_4$,}
\end{equation}
otherwise there would be a shorter directed cycle in $D(A)$.
Let $P$ be an induced $P_4$ containing $\{v_1,v_2\}$ as an end-edge. Let, 
without loss of generality, $P$ be $v_1v_2ab$ for some vertices $a$ and $b$. Since $(v_1,v_2)\in A$, $(b,a)\in B$, and hence $a\to b\in D(A)$. 

Note that, as $v_q$ and $v_1$ are adjacent, $a, b\not=v_q$. Moreover,
\begin{equation*}
\text{$v_q$ is non-adjacent to $b$.}
\end{equation*}
(Otherwise $G[v_q, v_1, v_2, a, b]$ would be a $C_5$, a house or a gem.)

If $q=3$, then $v_3v_2ab$ (if $a$ and $v_3$ are non-adjacent) or $v_1v_3ab$ (if $a$ and $v_3$ are adjacent) is a bad $P_4$, a contradiction.

If $q=4$, then $a$ and $v_4$ are non-adjacent (otherwise $v_1v_4ab$ would be a bad $P_4$). Then, $a\not= v_3$. Moreover, $v_2$ and $v_4$ are non-adjacent (otherwise $v_2v_4$ would be a chord of $C$ that is an end-edge of the induced $P_4\, v_4v_2ab$, contradicting (\ref{eq:coalition})), and $a$ and $v_3$ are non-adjacent (otherwise $C+a$ would be a house or a gem). But then $v_4v_1v_2a$ or $v_4v_3v_2a$ is a bad $P_4$, a contradiction.

So, let us assume that $q\ge 5$. Then, as $G$ is hole-free, $C$ has a chord connecting $v_i$ and $v_{i+2}$ (a {\it $3$-chord\/}) or $v_i$ and $v_{i+3}$ (a {\it $4$-chord\/}) for some $i$ (modulo $q$). We distinguish two cases.

\smallskip
\noindent \textsc{Case 1:} $C$ has a $3$-chord. We may assume that $\{v_q, v_2\}$ is a $3$-chord of $C$. Then $v_q$ and $a$ are adjacent (otherwise the chord $\{v_q, v_2\}$ of $C$ would be an end-edge of the induced $P_4\, v_qv_2ab$, contradicting (\ref{eq:coalition})). But now $v_1, v_q, a$ and $b$ induce a bad $P_4$. Case~1 is setteld.

\smallskip
\noindent \textsc{Case 2:} $C$ has no $3$-chord. In this case, we may assume that $\{v_q, v_3\}$ is a $4$-chord of $C$ (and notice that $v_i$ and $v_{i+2}$ are non-adjacent for all $i$). Then $v_q$ and $a$ are non-adjacent (otherwise $v_1, v_q, a$ and $b$ would induce a bad $P_4$). Then, $a\not=v_3$. Moreover, $a$ and $v_3$ are non-adjacent (otherwise $G[v_q, v_1,v_2, v_3,a]$ would be a house). But then the chord $\{v_q, v_3\}$ of $C$ would be an end-edge of the induced $P_4\, v_qv_3v_2a$, contradicting (\ref{eq:coalition}). Case~2 is settled.

\smallskip
The proof of Theorem~\ref{thm:gemhouseholecoali} is complete.\qed

\begin{corollary}
Given a $(\text{gem, house, hole})$-free graph $G$ with $m$ edges, it can be recognized in $O(m^{2})$ time if $G$ is a coalition graph, and if so, a coalition orientation of $G$ can be obtained in the same time.
\end{corollary}

For the smaller class of distance-hereditary graphs, we can do better.
Note that the distance-hereditary graph $\mathsf{N}$ depicted in Figure~\ref{fig:N} is a minimal non-coalition graph.

\begin{theorem}\label{thm:dhcoali}
The following statements are equivalent for a distance-hereditary graph $G$:
\begin{itemize}
 \item[\em (i)] $G$ is a coalition graph;
 \item[\em (ii)] $\mathscr{C}(G)$ is bipartite;
 \item[\em (iii)] $G$ is $\mathsf{N}$-free $(\text{see Figure}~\ref{fig:N})$;
 \item[\em (iv)] $G$ is a comparability graph.
\end{itemize}
\end{theorem}
\proof (i) $\Rightarrow$ (ii) follows from Theorem~\ref{thm:gemhouseholecoali}.
The implications (ii) $\Rightarrow$ (iii) and (iv) $\Rightarrow$ (i) are obvious (recall that comparability graphs are coalition graphs). 
The implication (iii) $\Rightarrow$ (iv) has been proved in~\cite[Theorem 3.1]{DiStefano}.

We remark that (iii) $\Rightarrow$ (iv) can also be proved by induction on the number of vertices of $G$ as follows. Let $G$ be an $\mathsf{N}$-free distance-hereditary graph. Then $G$ has a pair of twins or a pendant vertex (\cite{BM}).
If $x, y$ are twins, then by induction, $G-x$ admits a transitive orientation. Adopting the directions of the $y$-edges for the $x$-edge and orienting the edge $xy$ (if any) arbitrarily, we obtain a transitive orientation for $G$. 
So, let $G$ have a pendant vertex $u$ and let $v$ be the neighbor of $u$. By induction again, $G-u$ is a comparability graph. As $G$ has no gem, no house, and no $\mathsf{N}$, $v$ is a \lq regular\rq\ vertex of $G-u$ in the sense of \cite{Olariu1992}. It follows from~\cite[Theorem 1]{Olariu1992} that $G-u$ admits a transitive orientation in which $v$ is a source. Then orienting $uv$ from $v$ to $u$ we obtain a transitive orientation for $G$.
Hence (iv). \qed

\medskip
Theorem~\ref{thm:dhcoali} implies that distance-hereditary coalition graphs can be recognized in time needed to recognize comparability graphs which requires time proportional to matrix multiplication (see \cite{BLS,Spinrad} for more information). However, \cite[Theorem 4.14]{DiStefano} provides linear time recognition algorithm for distance-hereditary comparability graphs. Thus, we have    

\begin{corollary}
Given a distance-hereditary graph $G$ with $n$ vertices and $m$ edges, it can be recognized in $O(n+m)$ time if $G$ is a coalition graph, and if so, a coalition orientation of $G$ can be obtained in the same time.
\end{corollary}

\begin{corollary}
A distance-hereditary graph is a comparability graph if and only if it is $\mathsf{N}$-free.
\end{corollary}

\section{Concluding remarks}
Recognizing and characterizing coalition graphs and opposition gra\-phs remain long-standing open problems in structural and algorithmic graph theory.
In this paper we have characterized generalized opposition graphs, and our characterization leads to a polynomial time recognition algorithm. We also give algorithmic good characterizations, as well as forbidden subgraph characterizations for some opposition graph classes and coalition graph classes, including distance-hereditary opposition and coalition graphs.

We conclude the paper with some obvious problems for future research:

\begin{enumerate}
 \item Determine further (perfect) graph classes $\mathcal{G}$ such that opposition graphs and generalized opposition graphs in $\mathcal{G}$ coincide, i.e., for all $G\in\mathcal{G}$, $G$ is an opposition graph if and only if $\mathscr{O}(G)$ is bipartite. For such classes, opposition graphs can be recognized in polynomial time. This is done for $\mathcal{G}=\text{bipartite graphs}$ in~\cite{Le2013} and for  $\mathcal{G}=\text{distance-hereditary graphs}$ in this paper.

 \item Similarly, we know that if $G$ is a coalition graph, then $\mathscr{C}(G)$ is bipartite. For which graph classes does the converse hold true? In this paper, it is shown that this is the case for distance-hereditary graphs.

 \item Characterize and recognize chordal opposition/coalition graphs. In this paper we are able to solve the case of ptolemaic graphs.

 \item Characterize and recognize comparability graphs that are opposition graphs,
  and co-comparability graphs that are opposition/coalition gra\-p\-hs.
  In~\cite{Le2013}, the case of bipartite graphs and co-bipartite graphs are settled.

\end{enumerate}

Finally, it would be interesting to know if there is an opposition graph that is not perfectly orderable. Olariu~\cite{Olariu1997} conjectures that such an opposition graph exists.

\medskip\noindent
\textbf{Acknowledgments.}\, We thank the two unknown referees for their very careful reading and helpful remarks that led to improvements in the presentation of the paper.


\begin{thebibliography}{99}
\bibitem{BM}
  Hans-J\"{u}rgen Bandelt, Henry M. Mulder,
  Distance-hereditary graphs,
  {\it J. Combinatorial Theory\/} Series B 41 (1986) 182--208.

\bibitem{BLS}
  Andreas Brandst\"adt, Van Bang Le, Jeremy P. Spinrad,
  {\it Graph Classes: A Survey\/}, SIAM Monographs on Discrete Math. Appl.,
   Vol. 3, Philadelphia, 1999.

\bibitem{CRST}
  Maria Chudnovsky, Neil Robertson, Paul Seymour, Robin Thomas,
  The strong perfect graph theorem,
  {\it Annals of Mathematics\/} 164 (2006) 51--229.

\bibitem{Chvatal1984}
  Va{\v s}ek Chv\'atal,
  Perfectly ordered graphs,
  In: Topics on Perfect Graphs (C. Berge, V. Chv\'atal, eds.),
  {\it Annals of Discrete Mathematics\/} 21 (1984) 63--65.

\bibitem{Chvatal1993}
  Va{\v s}ek Chv\'atal,
  Which claw-free graphs are perfectly orderable?
  {\it Discrete Applied Mathematics\/} 44 (1993) 39--63.

\bibitem{Chvatal2000}
  Va{\v s}ek Chv\'atal,
  Generalized opposition graphs,
  In: {\em Open problems on perfect graphs\/}, 1.~Perfection of special classes of Berge graphs. 
  http:/\!/www.cs.rutgers.edu/\~{}chvatal/perfect/problems.html,
  August 24, 2000.

\bibitem{DiStefano}
  Gabriele Di Stefano, 
  Distance-hereditary comparability graphs,
  {\it Discrete Applied Mathematics\/} 160 (2012) 2669--2680.

\bibitem{EJSS}
  Elaine M. Eschen, Julie L. Johnson, Jeremy P. Spinrad, R. Sritharan,
  Recognition of some perfectly orderable graph classes,
  {\it Discrete Applied Mathematics\/} 128 (2003) 355--373.

\bibitem{Hertz}
  Alain Hertz,
  Bipolarizable graphs,
  {\it Discrete Mathematics\/} 81 (1990) 25--32.

\bibitem{Hoang1996b}
  Ch\'{\i}nh T. Ho\`ang,
  On the complexity of recognizing a class of perfectly orderable graphs,
  {\it Discrete Applied Mathematics\/} 66 (1996) 219--226.

\bibitem{Hoang2001}
  Ch\'{\i}nh T. Ho\`ang,
  Perfectly Orderable Graphs: A Survey,
  in: J.L. Ram\'{\i}rez Alfons\'{\i}n, B.A. Reed (eds.), \emph{Perfect Graphs\/},
  Wiley Interscience, New York, 2001.

\bibitem{HoangReed}
  Ch\'{\i}nh T. Ho\`ang, Bruce A. Reed,
  Some classes of perfectly orderable graphs,
  {\it J. Graph Theory} 13 (1989) 445--463.

\bibitem{Le2013}
  Van Bang Le, On opposition graphs, coalition graphs, and bipartite permutation graphs,
  {\it Discrete Applied Mathematics\/} 168 (2014) 26--33.


\bibitem{LinOla}
  D. Link, Stefan Olariu,
  A simple NC algorithm for Welsh-Powell opposition graphs,
  {\it Intern. J. Computer Math.\/}, 41 (1991) 49--53.

\bibitem{MidPfe}
  Matthias Middendorf, Frank Pfeiffer,
  On the complexity of recognizing perfectly orderable graphs,
  {\it Discrete Mathematics\/} 80 (1990) 327--333.

\bibitem{NP}
  Stavros D. Nikolopoulos, Leonidas Palios,
  On the Recognition of Bipolarizable and $P_4$-simplicial Graphs,
  {\it Discrete Math. \& Theoretical Computer Science\/}, 7 (2005) 231--254.

\bibitem{NikPal}
  Stavros D. Nikolopoulos, Leonidas Palios,
  Recognizing HH-free, HHD-free, and Welsh-Powell Opposition Graphs,
  {\it Discrete Math. \& Theoretical Computer Science\/}, 8 (2006) 65--82.  
  
\bibitem{Olariu}
  Stephan Olariu,
  All Variations on Perfectly Orderable Graphs,
  {\it J. Combin. Theory\/}, Series B 45 (1988) 150--159.

\bibitem{Olariu1992}
  Stephan Olariu,
  On sources in comparability graphs, with applications,
  {\it Discrete Mathematics\/}, 110 (1992) 289--292.

\bibitem{Olariu1997}
  Stephan Olariu,
  Private communication, Berlin, Rostock, September 1997.

\bibitem{OlaRan}
  Stephan Olariu, J. Randall,
  Welsh-Powell opposition graphs,
  {\it Information Processing Letters\/}, 31 (1989) 43--46.
 
\bibitem{Spinrad}
  Jeremy P. Spinrad,
  {\it Efficient Graph Representations\/},
  Fields Institute Monographs Vol. 19, American Mathematical Society, Providence, 2003.

\end{thebibliography}
\end{document}